


\newif\ifPubSub \PubSubfalse
\newcount\PubSubMag \PubSubMag=1200
\def\PubSub{\PubSubtrue
            \magnification=\PubSubMag \hoffset=0pt \voffset=0pt
            \pretolerance=600 \tolerance=1200 \vbadness=1000 \hfuzz=2 true pt
            \baselineskip=1.75\baselineskip plus.2pt minus.1pt
            \parskip=2pt plus 6pt
            \setbox\strutbox=\hbox{\vrule height .75\baselineskip
                                               depth  .25\baselineskip
                                               width 0pt}%
            \Page{6 true in}{8.9 true in}}

\newcount\FigNo \FigNo=0
\newbox\CapBox
\newbox\FigBox
\newtoks\0
\def\Fig {fig.~\the\FigNo}
\def\NFig {{\count255=\FigNo \advance\count255 by 1 fig.~\the\count255}}
\def\StartFigure #1#2#3{\global\advance\FigNo by 1
                  \ifPubSub \global\setbox\FigBox=\vbox\bgroup
                            \unvbox\FigBox
                            \parindent=0pt \parskip0pt
                            \eject \line{\hfil}\bigskip\bigskip
                  \else \midinsert \removelastskip \bigskip\bigskip
                  \fi
                  \begingroup \hfuzz1in
                  \dimen0=\hsize \advance\dimen0 by-2\parindent \indent
                  \vbox\bgroup\hsize=\dimen0 \parindent=0pt \parskip=0pt
                       \vrule height1pt depth0pt width0pt
                       \ifdim\dimen0<#1bp \dimen1=#1bp
                                          \advance\dimen1 by -\dimen0
                                          \divide\dimen1 by 2 \hskip-\dimen1
                       \fi
                       \hfil
                       \vbox to #2 bp\bgroup\hsize #1 bp
                            \vss\noindent\strut\special{"#3}%
                            \skip0=\parskip \advance\skip0 by\dp\strutbox
                            \vskip-\skip0 }%
\def\caption #1{\strut\egroup
                \ifPubSub \global\setbox\CapBox=\vbox{\unvbox\CapBox
                                        \parindent0pt \medskip
                                         {\bf Figure \the\FigNo:}
                                          {\tenrm\strut #1\strut}}%
                \else \bigskip {\FootCapFace \hfuzz=1pt \baselineskip=2.6ex
                                \noindent{\bf Figure~\the\FigNo:} #1\par}%
                \fi}%
\def\label (#1,#2)#3{{\offinterlineskip \parindent=0pt \parskip=0pt
                     \vskip-\parskip 
                     \vbox to 0pt{\vss
                          \moveright #1bp\hbox to 0pt{\raise #2bp
                                         \hbox{#3}\hss}\hskip-#1bp\relax}}}%
\def\EndFigure {\egroup \endgroup 
                \ifPubSub \vfil
                          \centerline{\bf Figure \the\FigNo}
                          \egroup 
                \else \bigskip \smallskip \endinsert 
                \fi }
\def\ListCaptions {\vfil\eject \message{!Figure captions:!}%
                   \Sectionvar{Figure Captions}\par
                   \unvbox\CapBox}
\def\ShowFigures {\ifPubSub\ifnum\FigNo>0 \ListCaptions \vfil\eject
                                 \message{!Figures:!}%
                                 \nopagenumbers \unvbox\FigBox \eject
                           \fi
                  \fi}
\def\Sectionvar #1{\leftline{\bf #1}\bigskip}
\def\FootCapFace{} 


\parskip=0pt plus 3pt
\def\Page #1#2{{\dimen0=\hsize \advance\dimen0 by-#1 \divide\dimen0 by 2
               \global\advance\hoffset by \dimen0
               \dimen0=\vsize \advance\dimen0 by-#2 \divide\dimen0 by 4
               \ifdim\dimen0<0pt \multiply\dimen0 by 3 \fi
               \global\advance\voffset by \dimen0
               \global\hsize=#1 \global\vsize=#2\relax}
               \ifdim\hsize<5.5in \tolerance=300 \pretolerance=300 \fi}
\Page{5in}{8in} 
\def\EndPaper{\par\dosupereject \ifnum\RefNo>1 \ShowReferences \fi
              \ShowFigures
              \par\vfill\supereject
              \message{!!That's it.}\end}
\headline{\ifnum\pageno=-1 \hfil \Smallrm \Time, \Date \else \hfil \fi}
\def\VersionInfo #1{\headline{\ifnum\pageno=-1 \hfil \Smallrm #1
                              \else \hfil
                              \fi}}


\newcount\RefNo \RefNo=1
\newbox\RefBox
\def\Jou #1{{\it #1}}
\def\Vol #1{{\bf #1}}
\def\AddRef #1{\setbox\RefBox=\vbox{\unvbox\RefBox
                       \parindent1.75em
                       \pretolerance=1000 \hbadness=1000
                       \vskip0pt plus1pt
                       \item{\the\RefNo.}
                       \sfcode`\.=1000 \strut#1\strut}%
               \global\advance\RefNo by1 }
\def\Ref  #1{\Hunskip~[{\the\RefNo}]\AddRef{#1}}
\def\Refc #1{\Hunskip~[{\the\RefNo,$\,$}\AddRef{#1}}
\def\Refm #1{\Hunskip{\the\RefNo,$\,$}\AddRef{#1}}
\def\Refe #1{\Hunskip{\the\RefNo}]\AddRef{#1}}
\def\Refl #1{\Hunskip~[{\the\RefNo}--\nobreak\AddRef{#1}}
\def\Refn #1{\Hunskip\AddRef{#1}}
\def\ShowReferences {\message{!References:!}%
                     \Sectionvar{References}\par
                     \unvbox\RefBox}
\def\StoreRef #1{\Hunskip\edef#1{\the\RefNo}}


\newif\ifRomanNum
\newcount \Sno \Sno=0
\def\Interskip #1#2#3{{\removelastskip \dimen0=#1
                       \advance\dimen0 by2\baselineskip
                       \vskip0pt plus\dimen0 \penalty-300
                       \vskip0pt plus-\dimen0
                       \advance\dimen0 by -2\baselineskip
                       \vskip\dimen0 plus#2 minus#3}}
\def\Section #1\par{\Interskip{24pt}{6pt}{2pt}%
                    \global\advance\Sno by1
                    \setbox0=\hbox{\STitlefont
                                    \ifRomanNum
                                       \global\SSno=64 \uppercase
                                       \expandafter{\romannumeral
                                                    \the\Sno.\ \ }%
                                     \else
                                        \global\SSno=96 \the\Sno.\ \
                                     \fi}
                    \leftline{\vtop{\copy0 }%
                              \vtop{\advance\hsize by -\wd0
                                    \raggedright
                                    \pretolerance10000 \hbadness10000
                                    \noindent \STitlefont
                                    \GetParenDim{\dimen0}{\dimen1}%
                                    \advance\dimen0 by\dimen1
                                    \baselineskip=\dimen0 #1}}
                    \hrule height0pt depth0pt
                    \dimen0=\baselineskip \advance\dimen0 by -\parskip
                    \nobreak\vskip\dimen0 plus 3pt minus3pt \noindent}%
\def\Sectionvar #1\par{\Interskip{24pt}{6pt}{2pt}%
                    \leftline{\vbox{\noindent
                                    \STitlefont #1}}
                    \hrule height0pt depth0pt
                    \dimen0=\baselineskip \advance\dimen0 by -\parskip
                    \nobreak\vskip\dimen0 plus 3pt minus3pt \noindent}%
\newcount\SSno \SSno=0
\def\SubSection #1\par{\Interskip{15pt}{3pt}{1pt}%
                    \global\advance\SSno by1
                    \setbox0=\hbox{\SSTitlefont \char\SSno$\,$]$\,$\ }
                    \leftline{\vtop{\copy0 }%
                              \vtop{\advance\hsize by -\wd0
                                    \raggedright
                                    \pretolerance10000 \hbadness10000
                                    \noindent \SSTitlefont
                                    \GetParenDim{\dimen0}{\dimen1}%
                                    \advance\dimen0 by\dimen1
                                    \baselineskip=\dimen0 #1}}
                    \hrule height0pt depth0pt
                    \dimen0=.8\baselineskip \advance\dimen0 by -2\parskip
                    \nobreak\vskip\dimen0 plus1pt minus1pt \noindent}%
\def\(#1){~\ifRomanNum {\Medrm\uppercase\expandafter{\romannumeral #1}}%
           \else {#1}%
           \fi}


\newcount\EqNo \EqNo=0
\def\NumbEq {\global\advance\EqNo by 1
             \eqno(\the\EqNo)}
\def\PrevEq {(\the\EqNo)}
\def\PrevEqs #1{{\count255=\EqNo \advance\count255 by-#1\relax
                 (\the\count255)}}
\def\NameEq #1{\xdef#1{(\the\EqNo)}}

\def\AppndEq #1{\EqNo=0
  \def\NumbEq {\global\advance\EqNo by 1
             \eqno(#1\the\EqNo)}
  \def\numbeq {\global\advance\EqNo by 1
             (#1\the\EqNo)}
  \def\PrevEq {(#1\the\EqNo)}
  \def\PrevEqs ##1{{\count255=\EqNo \advance\count255 by-##1\relax
                 (#1\the\count255)}}
  \def\NameEq ##1{\xdef##1{(#1\the\EqNo)}}}


\newcount\ThmNo \ThmNo=0

\def\Theorem #1\par{\removelastskip\bigbreak
    \advance\ThmNo by 1
    \noindent{\bf Theorem \the\ThmNo:} {\sl #1\bigskip}}
\def\Lemma #1\par{\removelastskip\bigbreak
    \advance\ThmNo by 1
    \noindent{\bf Lemma \the\ThmNo:} {\sl #1\bigskip}}
\def\ContinueThm{\par\noindent}
\def\PrevThm {Theorem~\the\ThmNo}
\def\Thm {{\advance\ThmNo by 1 \PrevThm}}
\def\PrevLm {lemma~\the\ThmNo}
\def\Declare #1#2\par{\removelastskip\bigbreak
    \noindent{\bf #1:} {\sl #2\bigskip}}
\def\Proof {\removelastskip\bigskip
    \noindent{\bf Proof:\ \thinspace}}
\def\EndProof{{~\nobreak\hfil \copy\Tombstone
               \parfillskip=0pt \bigskip}}


\newlinechar=`\!
\def\\{\ifhmode\hfil\break\fi}
\let\thsp=\,
\def\,{\ifmmode\thsp\else,\thinspace\fi}
\def\Hunskip {\ifhmode\unskip\fi}

\def\Date {\ifcase\month\or January\or February\or March\or April\or
 May\or June\or July\or August\or September\or October\or November\or
 December\fi\ \number\day, \number\year}

\newcount\mins  \newcount\hours
\def\Time{\hours=\time \mins=\time
     \divide\hours by60 \multiply\hours by60 \advance\mins by-\hours
     \divide\hours by60         
     \ifnum\hours=12 12:\ifnum\mins<10 0\fi\number\mins~P.M.\else
       \ifnum\hours>12 \advance\hours by-12
         \number\hours:\ifnum\mins<10 0\fi\number\mins~P.M.\else
            \ifnum\hours=0 \hours=12 \fi
         \number\hours:\ifnum\mins<10 0\fi\number\mins~A.M.\fi
     \fi }

\def\HollowBox #1#2#3{{\dimen0=#1
       \advance\dimen0 by -#3 \dimen1=\dimen0 \advance\dimen1 by -#3
        \vrule height #1 depth #3 width #3
        \hskip -#3
        \vrule height 0pt depth #3 width #2
        \llap{\vrule height #1 depth -\dimen0 width #2}%
       \hskip -#3
       \vrule height #1 depth #3 width #3}}
\newbox\Tombstone
\setbox\Tombstone=\vbox{\hbox{\HollowBox{8pt}{5pt}{.8pt}}}

\def\GetParenDim #1#2{\setbox1=\hbox{(}%
                      #1=\ht1 \advance#1 by 1pt
                      #2=\dp1 \advance#2 by 1pt}

\def\Bull #1#2\par{\removelastskip\smallbreak\textindent{$\bullet$}
                    {\it #1}#2\smallskip}
\def\Heading #1{\removelastskip\smallbreak\noindent {\it #1.}%
                \nobreak\par\noindent\ignorespaces}


\def\Title #1\par{\message{!!#1}%
                  \nopagenumbers \pageno=-1
                  \null \bigskip
                  {\leftskip=0pt plus 1fil \rightskip=\leftskip
                   \parfillskip=0pt\relax \Titlefont
                   \ifdim\baselineskip<2.5ex \baselineskip=2.5ex \fi
                   \noindent #1\par}\bigskip}
\def\Author #1\par{{\bigskip
                     \count1=0   
                     \count2=0   
                     \dimen0=0pt 
                    \Position{#1\hskip-\dimen0 }\bigskip}}
\def\Address #1\par{{\leftskip=\parindent plus 1fil \rightskip=\leftskip
                     \parfillskip=0pt\relax
                     \ifdim\baselineskip>3ex \baselineskip=3ex \fi
                     #1\par}\bigskip}
\def\modfnote #1#2{{\parindent=1.1em \leftskip=0pt \rightskip=0pt
                    \GetParenDim{\dimen1}{\dimen2}%
                    \setbox0=\hbox{\vrule height\dimen1 depth\dimen2
                                          width 0pt}%
                    \advance\dimen1 by \dimen2 \baselineskip=\dimen1
                    \vfootnote{#1}{\hangindent=\parindent \hangafter=1
                                 \unhcopy0 #2\unhbox0
                                 \vskip-\baselineskip \vskip1pt}}}%
\def\PAddress #1{\ifcase\count1 \let\symbol=\dag
                 \or \let\symbol=\ddag
                 \or \let\symbol=\P
                 \or \let\symbol=\S
                 \else \advance\count1 by -3
                       \def\symbol{\dag_{\the\count1 }}%
                 \fi
                 \advance\count1 by 1
                 \setbox0=\hbox{$^{\symbol}$}\advance\dimen0 by \wd0
                 \Hunskip \box0
                 \modfnote{$\symbol$}{{\sl Permanent address\/}: #1}}%
\def\Email #1{\ifcase\count2 \let\symbol=\ast
                 \or \let\symbol=\star
                 \or \let\symbol=\bullet
                 \or \let\symbol=\diamond
                 \or \let\symbol=\circ
                 \else \advance\count2 by -4
                       \def\symbol{\ast_{\the\count2 }}%
                 \fi
                 \advance\count2 by 1
                 \setbox0=\hbox{$^{\textstyle\symbol}$}\advance\dimen0 by \wd0
                 \Hunskip \box0
                 \modfnote{$\symbol$}{{\sl Electronic mail\/}: #1}}%
\def\Abstract{\vfil \message{!Abstract:!}%
              \Position{{\STitlefont Abstract}}
              \medskip \bgroup
              \ifdim\baselineskip>3.5ex \baselineskip=3.5ex \fi
              \narrower\noindent\ignorespaces}%
\def\EndAbstract{\par\egroup}
\def\pacs #1{\vfil\leftline{PACS numbers: #1}\eject}
\def\StartPaper{\message{!Body:!}%
                \pageno=1 \ifPubSub \footline{\tenrm\hss
                                              --\ \folio\ -- \hss}%
                          \else \footline{\hss\vtop to 0pt{\hsize=.15\hsize
                                     \vglue6pt \hrule \medskip
                                     \centerline{\tenrm\folio}\vss}\hss}%
                          \fi}


\RomanNumtrue
\let\Position=\centerline 

\def\Titlefont{\tenbf}   
\def\STitlefont{\tenbf}  
\def\SSTitlefont{\tensl} 
\def\Smallrm{\sevenrm}   
\def\Medrm{\tenrm}       

\newcount\EZReadMag \EZReadMag=1200
\def\EZRead{\magnification=\EZReadMag \hoffset=0pt \voffset=0pt
            \pretolerance=1000 \tolerance=2000
            \vbadness=5000 \hbadness=1500 \hfuzz=3 true pt
            \baselineskip=1.1\baselineskip plus.1pt minus0pt
            \parskip=2pt plus 2pt
            \setbox\strutbox=\hbox{\vrule height .75\baselineskip
                                               depth  .25\baselineskip
                                               width 0pt}%
            \ifnum\EZReadMag=1050 \Page{5.75 true in}{8.75 true in}%
            \else \Page{5.875 true in}{8.75 truein}%
            \fi}

\message{!REMINDER:!
         If you print the dvi file via `dvips' the figures will!
         automatically emerge. If you use anything else, you will!
         get the text of the paper, but no figures.!
         Send e-mail to borde@bnlcl6.bnl.gov if you have any problems.!!}

\EZRead
\VersionInfo{June 30, 1994}

\Title TOPOLOGY CHANGE IN CLASSICAL GENERAL RELATIVITY

\Author Arvind Borde
\PAddress{Long Island University, Southampton, NY 11968, and\\
          High Energy Theory Group, Brookhaven National Laboratory,
          Upton, NY 11973.}
\Email{borde@bnlcl6.bnl.gov}

\Address Institute of Cosmology, Department of Physics and Astronomy,\\
         Tufts University, Medford, MA 02155

\Abstract
This paper clarifies some aspects of Lorentzian topology change, and it
extends to a wider class of spacetimes previous results of Geroch and Tipler
that show that topology change is only to be had at a price. The scenarios
studied here are ones in which an initial spacelike surface is joined by a
connected ``interpolating spacetime'' to a final spacelike surface, possibly
of different topology. The interpolating spacetime is required to obey a
condition called {\it causal compactness}, a condition satisfied in a very wide
range of situations. No assumption is made about the dimension of spacetime.
First, it is stressed that topology change is kinematically possible; i.e., if
a field equation is not imposed, it is possible to construct topology-changing
spacetimes with non-singular Lorentz metrics. Simple 2-dimensional examples of
this are shown. Next, it is shown that there are problems in such spacetimes:
Geroch's closed-universe argument is applied to causally compact spacetimes to
show that even in this wider class of spacetimes there are causality
violations associated with topology change. It follows from this result that
there will be causality violations if the initial (or the final) surface is
not connected, even when there is no topology change. Further, it is shown
that in dimensions $\geq 3$ causally compact topology-changing spacetimes
cannot satisfy Einstein's equation (with a reasonable source); i.e., there
are severe dynamical obstructions to topology change. This result extends a
previous one due to Tipler. Like Tipler's result, it makes no assumptions
about geodesic completeness; i.e., it does not permit topology change even at
the price of singularities (of the standard incomplete-geodesic variety).
Brief discussions are also given of the restrictions that are placed on the
source in this result, of the possibility of creating exotic differentiable
manifolds, and of ways in which the results of this paper might be
circumvented.
\EndAbstract


\StartPaper

\Section Introduction

If space has a certain topology at some initial time, can the topology be
different at a later time? The question has, of course, animatedly been
discussed before by a number of people
\StoreRef{\WhOne}
\Refl{J. Wheeler, in \Jou{Relativity Groups and Topology}, edited by
B.S. Dewitt and C.M. DeWitt, Gordon and Breach, New York (1963).}
\StoreRef{\WhTwo}
\Refn{J. Wheeler, \Jou{Geometrodynamics}, Academic Press, New York (1962).}
\StoreRef{\Hawk}
\Refn{S.W. Hawking, \Jou{Nuclear Physics}, \Vol{B144}, 349 (1978);
in \Jou{General Relativity: An Einstein Centenary Survey}, edited by
S.W. Hawking and W. Israel, Cambridge University Press, Cambridge (1979).}
\StoreRef{\MisWh}
\Refn{C.W. Misner and J. Wheeler, \Jou{Ann. of Physics (NY)}, \Vol{2},
525 (1957).}
\StoreRef{\Gero}
\Refn{R.P. Geroch, \Jou{J. of Math. Phys.}, \Vol{8}, 782 (1967).}
\StoreRef{\GeroPhD}
\Refn{R.P. Geroch, \Jou{Singularities in the spacetime of General
Relativity}, Ph.D. Dissertation, Princeton University (1968).}
\StoreRef{\TiplPhD}
\Refn{F.J. Tipler, \Jou{Causality Violations in General Relativity},
Ph.D. Dissertation, Univ. of Maryland (1976).}
\StoreRef{\Tipl}
\Refn{F.J. Tipler, \Jou{Ann. of Physics (NY)}, \Vol{108}, 1 (1977).}
\Refn{F.J. Tipler, \Jou{Phys. Lett.}, \Vol{165B}, 67 (1985).}
\StoreRef{\FrSork}
\Refn{J.L. Friedman and R. Sorkin, \Jou{Phys. Rev. Lett.}, \Vol{44}, 1100,
(1980).}
\Refn{A. Strominger, \Jou{Phys. Rev. Lett.}, \Vol{52}, 1733 (1984).}
\StoreRef{\SorkOne}
\Refn{R. Sorkin, in \Jou{Topological Properties and Global Structures of
spacetime}, edited by P.G. Bergmann and V. de Sabhata, Plenum, New York
(1986).}
\StoreRef{\SorkTwo}
\Refn{R. Sorkin, \Jou{Phys. Rev.}, \Vol{D33}, 978 (1986).}
\StoreRef{\ADeW}
\Refn{A. Anderson and B. DeWitt, \Jou{Found. of Physics}, \Vol{16}, 91
(1986).}
\StoreRef{\BGR}
\Refn{F.A. Bais, C. Gomez and V.A. Rubakov, \Jou{Nucl. Phys.},
\Vol{B282}, 531 (1987).}
\StoreRef{\MCD}
\Refn{C.A. Manogue, E. Copeland and T. Dray, \Jou{Pramana}, \Vol{30}, 279
(1988).}
\StoreRef{\MTY}
\Refn{M.S. Morris, K.S. Thorne and U. Yurtsever, \Jou{Phys. Rev. Lett.},
\Vol{61}, 1446 (1988).}
\Refn{M. Visser, \Jou{Phys. Rev.}, \Vol{D41}, 1116 (1990).}
\Refn{J. Friedman, p.~539 in \Jou{Conceptual Problems of Quantum Gravity},
edited by A.~Ashtekar and J.~Stachel, Birkhauser, Boston (1991).}
\StoreRef{\Horow}
\Refn{G. Horowitz, \Jou{Class. Quant. Grav.}, \Vol{8}, 587 (1991);
p.~1167 in the \Jou{Proceedings of the Sixth Marcel
Grossmann Meeting (Kyoto, Japan)}, World Scientific, Singapore (1992).}
\StoreRef{\GibHawk}
\Refn{G. Gibbons and S.W. Hawking, \Jou{Comm. in Math Phys.}, \Vol{148},
345 (1992); \Jou{Phys. Rev. Lett.}, \Vol{69}, 1719 (1992).}
\Refn{J. Friedman, K. Schleich and D. Witt,
\Jou{Phys. Rev. Lett.}, \Vol{71}, 1486 (1993).}
\Refe{A. Vilenkin, Tufts Institute of Cosmology preprint (1993).},
mostly in the context
of quantum gravity. In a quantum theory of gravitation, it has been
argued~[\WhOne--\Hawk], we will necessarily have to consider fluctuations
not only in geometry but also in topology. Topology change is also interesting
for another reason. It has sometimes been suggested~[\WhTwo\,\MisWh\,\SorkOne]
that the particles of ordinary matter might possibly be viewed
as kinks or knots in space. In this approach, nontrivial topological
configurations of space (sometimes called {\it geons\/}) are shown to
display such particle-like aspects as mass and charge~[\MisWh] and even
half-integral spin~[\FrSork]. Such theories would describe the creation
and annihilation of particles by allowing the spatial topology to change.
And finally, purely as a question about the classical Einstein theory,
we may ask: since geometry evolves in general relativity, might it not be
possible that topology does as well?

Topology change is interesting, therefore, for a number of somewhat
separate reasons. There is, however, a widespread notion that the
process is intrinsically incompatible with a Lorentzian metric. This notion
is quite incorrect. But, although Lorentzian topology change is possible,
it is problematic: it has been known for some time that
both in closed universes and in certain open ones there are problems
associated with topology change.
Geroch~[\Gero\,\GeroPhD] has shown that topology change may
be obtained in these cases only at the price of causality violations, and
Tipler~[\TiplPhD\,\Tipl] has shown that Einstein's equation cannot hold
(with a source with non-negative energy density) on such spacetimes if the
spatial topology changes. Part of the point of this paper is to
extend the results of Geroch and Tipler to a wider class of spacetimes,
and part of the point is to address some of the more widespread
misconceptions
about topology change. (These misconceptions are listed in section~IX.) Along
the way a few minor novelties~-- such as an explicit example of
non-singular 2-dimensional topology change~-- will be added to the general
topology-change discourse.

\SubSection A mechanism for topology change: classical evolution

To study topology change we first have to pick a mechanism through which
the topology might change. The most obvious choice is
``classical evolution;'' i.e., a scheme in which the initial
and final spatial configurations, represented respectively by
hypersurfaces ${\cal S}_1$ and ${\cal S}_2$, are joined by a
connected interpolating manifold $\cal M$ with a Lorentzian metric defined
on it (with respect to which~${\cal S}_1$ and~${\cal S}_2$ are spacelike).

Now, it may be tempting to argue, especially with the keen hindsight
provided by
the Geroch and Tipler results, that classical evolution is a patently
flawed mechanism~-- particularly if ones chief interest is quantum gravity.
But, for at least one approach to quantum gravity~-- the
path integral approach~-- the question of finding ``classical paths''
between the initial and final configurations is an important one.
(The paths discussed here are, however, Lorentzian ones, and
so the results of this paper are probably irrelevant to the Euclidean
path integral approach to quantum gravity~[\Hawk].)
The existence of classical paths is also important for work on
geons~[\SorkTwo], although here, too, Euclidean paths have been
considered~[\BGR]. And if we are interested in
topology change in the classical Einstein theory, then classical evolution
is (by definition) the mechanism to consider.

The appropriateness of classical evolution may be questioned at
another level. If the interpolating spacetime~$\cal M$ can be
foliated (i.e., sliced) into a family of spacelike hypersurfaces
(which corresponds to our immediate intuitive notion of an initial
spatial configuration evolving into some final configuration), then
it seems obvious that if the topology is to change,
$\cal M$ must contain cuts or holes of some kind (or some type
of singularity), and therefore be an unacceptable model for spacetime.
This statement can indeed be made precise and proved (this is done in
section~IV, in a corollary to theorem~1), but it still does not rule
out classical evolution as a mechanism.

There are two ways in which classical evolution may be rescued.
One way is to allow the metric to
become degenerate at isolated points~[\SorkTwo\,\Horow\,
\StoreRef{\Asht}
\Refe{A. Ashtekar, \Jou{Lectures on Non-perturbative Canonical Gravity},
World Scientific, Singapore (1991).},
while still retaining the ability to foliate~$\cal M$.
This approach will be briefly mentioned again in section~IX.
For the bulk of this paper, however, another way out is taken. Here, the
metric is required to behave itself everywhere, but the ability to foliate
is given up. This approach is based on the point of view that there is
no compelling reason why the spacetime that interpolates between the initial
and final states (which are the ones of physical interest) must be
restricted to be of the type that admits a foliation. Indeed, there
are solutions of Einstein's equation that do not admit a foliation into
spacelike hypersurfaces anywhere (e.g., the G\"odel solution), or which admit
foliations in some regions but not in others (e.g., the Taub-NUT
solution)%
\StoreRef{\HE}%
\Ref{S.W. Hawking and G.F.R Ellis, \Jou{The Large Scale Structure of
spacetime}, Cambridge Univ. Press, Cambridge (1973).}.
Once we allow these kinds of
interpolating spacetimes, then at the ``kinematical''
level (i.e., before imposing Einstein's equation) it is easy to
construct~-- as we shall see below~-- nonsingular topology-changing
spacetimes.

\SubSection An overview of the paper

After some preliminary comments in section~II,
the kinematics of topology change is discussed
in sections~III and~IV. A class of spacetimes,
called {\it causally compact\/} spacetimes, is defined in section~III and is
shown to cover many cases of physical interest. It is also argued that some
restriction on the interpolating spacetime, along the lines of causal
compactness, is indeed necessary. It is demonstrated that many
topology-changing causally compact spacetimes exist, and some
2-dimensional examples are shown. Then, in section~IV, a direct extension of
Geroch's theorem~[\Gero\,\GeroPhD] is presented which shows that such
topology-changing spacetimes must contain closed timelike curves.
Further, if causality violations are excluded,
then the initial and final surfaces must each be connected.
In section~V, the dynamics of topology change is
discussed and an extension of Tipler's theorem~[\TiplPhD\,\Tipl] is
presented that shows that the process of topology change cannot be described
by Einstein's equation (even if the spacetime has incomplete
geodesics). In this result certain
assumptions are made about the curvature of spacetime. The most stringent
of these
assumptions is discussed in section~VI, along with some reasonable
conditions on the source in Einstein's equation (the ``energy conditions'')
that will justify it. Upto this point in the paper most of the
discussion has assumed that the metric is time-orientable (i.e, that
the future can be globally distinguished from the past). In section~VII
it is shown that dropping this assumption does not materially alter
the conclusions of the main theorems (theorems~1 and~3) of this paper.
A few mathematical comments are made in section~VIII,
and some concluding remarks in section~IX.

\Section Notation, Definitions and other Preliminaries

The conventions and notation that I use are those of
Hawking and Ellis~[\HE]; the proofs of all the assertions that I make
below may be found there (unless otherwise indicated), as can further
references to the original statements and proofs of these assertions.

A {\it Lorentzian metric\/} on an $n$-dimensional manifold is a
metric with signature $(-,+,\ldots,+)$. Einstein's equation is
$$
R_{ab} -{1\over 2}Rg_{ab} + \Lambda g_{ab} = 8\pi T_{ab},
$$
where $R_{ab}$ is the Ricci tensor obtained from the metric~$g_{ab}$,
$R$~the
curvature scalar, $\Lambda$~the cosmological constant, and $T_{ab}$
the matter energy-momentum tensor. Einstein's equation is used
in a very minor way in this paper, only to justify the
assumptions of theorem~3. It will not be necessary to make any assumptions
about whether or not the cosmological constant~$\Lambda$ is zero.

The manifolds I consider are smooth, Hausdorff and paracompact.
A smooth manifold is one that is, essentially, infinitely
differentiable (i.e., $C^\infty $; a brief discussion of this is given in
section~VIII). The other two restrictions are technical ones; they ensure
that our models for spacetime are mathematically well-behaved.
These two restrictions will not appear explicitly in the rest of this paper.
On such manifolds it is always possible to define a smooth
positive-definite metric; this facility will be used below.
The manifolds are assumed to be
without boundary, unless explicitly stated otherwise.

\SubSection Interpolating spacetimes and time-orientation

When studying topology change, the region of interest is the spacetime
strip between an initial surface and a final surface.
This strip is usually part of some ``full spacetime,'' but it will
not be necessary at any stage for us to refer to this larger
spacetime~-- i.e., we
are only interested in what happens between the initial and final states,
not in what has happened before or is to happen after.
I consider in this paper situations that involve (roughly speaking)
either a ``fully spatially extended''
strip of spacetime or a portion of it confined to a
timelike tube. (See \NFig.)
This rough idea is captured in the following definitions:

Let ${\cal S}_1$ and ${\cal S}_2$ be disjoint
($n-1$)-dimensional manifolds, possibly with boundary.
A connected $n$-dimensional manifold~$\cal M$,
with boundary~$\partial \cal M$, is called an
{\it interpolating manifold\/} between ${\cal S}_1$ and ${\cal S}_2$, if
there is a further ($n-1$)-dimensional manifold~$\cal T$ (possibly empty)
such that~$\cal T$ is compact and
$\partial{\cal M}={\cal S}_1\cup{\cal S}_2\cup{\cal T}$.
I.e., the boundary of $\cal M$ is allowed, in addition to~${\cal S}_1$
and~${\cal S}_2$, to have an extra (compact) component~$\cal T$.
This component may be empty, but if it is not, then
it is required that
(i)~${\cal T}\cap {\cal S}_i = \partial{\cal S}_i\ne \emptyset$ ($i=1,2$),
(ii)~$\partial{\cal S}_1$ and $\partial{\cal S}_2$ have the same
topology (more precisely, are diffeomorphic~-- see section~VIII for a
definition of the term), and
(iii)~${\cal T}$ is diffeomorphic to $\partial{\cal S}_1\times [0,\,1]$.
The role played by~$\cal T$ will be seen shortly. Though $\cal M$ is
required to be connected, ${\cal S}_1$ and ${\cal S}_2$
are not restricted in this way.

The interpolating manifold $\cal M$ is called an
{\it interpolating spacetime\/} between ${\cal S}_1$ and ${\cal S}_2$,
if, in addition, there is a smooth Lorentz metric on it
with respect to which
${\cal S}_1$ and ${\cal S}_2$ are spacelike and $\cal T$, if non-empty,
is timelike in this sense: it is possible to define a smooth, nowhere
vanishing, timelike vector field everywhere tangent to~$\cal T$ such that
each integral curve of this field
takes on one endpoint at ${\cal S}_1$ and another at~${\cal S}_2$.
(Integral curves of a vector field are, roughly, curves to which the
field
is tangent; a more precise definition is given in the proof of \Thm.)
If the surface~$\cal T$ is non-empty, it represents a timelike tube within
which the topology change, if any, must occur. If~$\cal T$ is empty, then
it means that we are looking at a ``fully spatially extended''
strip of spacetime. The two scenarios are illustrated in~\NFig.
\StartFigure{356}{164}
        {0 20 translate
         /TopSurf {newpath 0 108 moveto
         40 118 100 98 144 108 curveto
         174 138 lineto
         144 130 75 144 30 138 curveto
         closepath } def
         /BotSurf {newpath 0 5 moveto
         48 0 105 12 144 5 curveto
         174 35 lineto
         138 42 80 28 30 35 curveto
         closepath }def
         TopSurf gsave .8 setgray fill grestore
         BotSurf gsave .8 setgray fill grestore
         190 0 translate
         BotSurf stroke
         gsave
         1 .5 scale
         newpath 74 36 15 0 360 arc .8 setgray fill
         grestore
         gsave [3 3] 0 setdash
         newpath 59 18 moveto 59 126 lineto stroke
         newpath 89 18 moveto 89 126 lineto stroke
         grestore
         TopSurf gsave 1 setgray fill grestore
         TopSurf stroke
         gsave
         1 .5 scale
         newpath 74 252 15 0 360 arc .8 setgray fill
         grestore
        }
\label(75,0){(a)}
\label(240,0){(b)}
\label(0,17){${\cal S}_1$}
\label(0,120){${\cal S}_2$}
\label(70,80){$\cal M$}
\label(280,35){${\cal S}_1$}
\label(280,138){${\cal S}_2$}
\label(259,80){$\cal M$}
\label(283,90){$\cal T$}
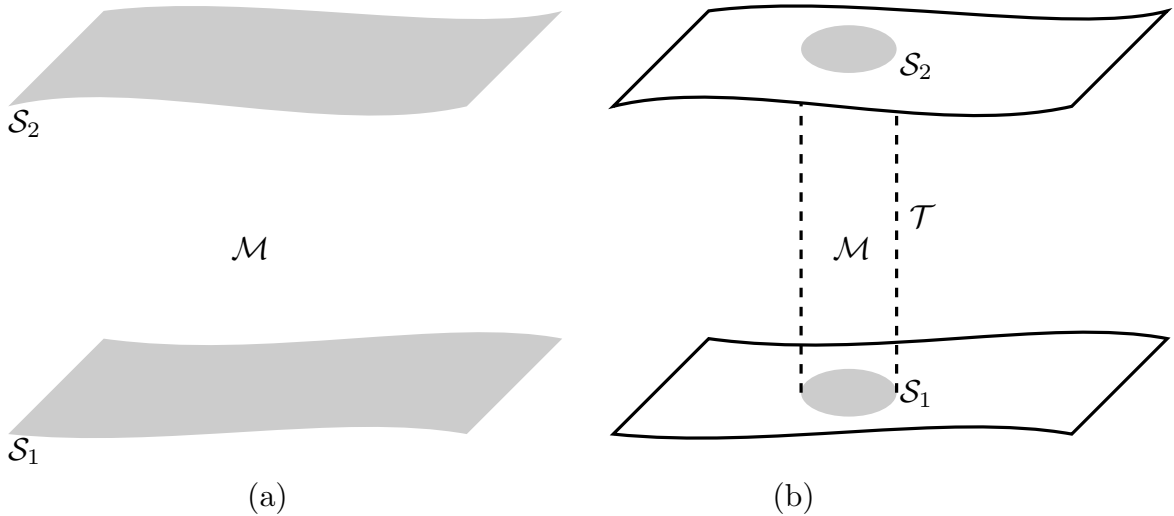
\caption{The two types of scenarios studied in this paper: (a)~a
fully spatially extended
spacetime strip, and (b)~a portion of such a strip confined to a
timelike tube. In each case, the shaded regions ${\cal S}_1$ and
${\cal S}_2$ represent, respectively, the initial and final surfaces, and
$\cal M$ the interpolating spacetime. In~(b), $\cal T$ represents
the timelike tube within which the topology change, if any, must occur.}
\EndFigure

The surface ${\cal S}_1$ will be called
an {\it initial surface\/} for the interpolating spacetime~$\cal M$ if no
$p\in \cal M$ lies to the past of a point on ${\cal S}_1$, and ${\cal S}_2$
will be called a {\it final surface\/} if no $p\in \cal M$ lies to
the future of a point on ${\cal S}_2$. It is possible that either of
${\cal S}_1$ or ${\cal S}_2$ may be empty: these cases will correspond,
respectively, to the creation and the annihilation of the universe.
(If they are both empty, $\cal M$ may be said to describe a situation in
which nothing comes from nothing.)

These definitions of initial and final surfaces assume that
it is possible to
consistently distinguish future from past throughout $\cal M$. A metric that
allows such a global choice of future and past is called
{\it time-orientable\/} (locally any Lorentz
metric permits such a choice: see ref.~[\HE], p.~38--39).
If such a choice of future direction
has actually been made, the metric is called {\it time-oriented}.
It is assumed that the metrics being considered are time-oriented.
No assumptions are made directly about the orientability
of the underlying manifold. (A non-orientable manifold like a M\"obius strip
may admit a time-orientable metric,
and an orientable manifold like an annulus
may admit a metric that is not time-orientable
\StoreRef{\GH}
\Ref{R.P. Geroch and G.T. Horowitz, in \Jou{General Relativity: an
Einstein Centenary Survey}, edited by S.W. Hawking and G.F.R. Ellis,
Cambridge University Press, Cambridge (1979).}).
If a result does depend on the orientability of the manifold, I
will explicitly point this out. Questions of orientability are
discussed again a little in section~VII,
especially the question of non-time-orientable metrics.

\SubSection Lorentz metrics and vector fields

The existence of a time-orientable, smooth Lorentz metric $g_{ab}$
on a manifold $\cal M$ (with or without boundary) is equivalent to the
existence of a smooth, nowhere vanishing vector field, $V^a$,
on $\cal M$ (ref.~[\HE], p.~39 and p.~181). To see this, first choose a
positive-definite metric $h_{ab}$ on~$\cal M$. Given~$V^a$,
define $g_{ab} = (h_{cd}V^cV^d)h_{ab} - 2h_{ac}V^ch_{bd}V^d$. This
$g_{ab}$ will be a Lorentz metric and $V^a$ will be timelike with respect
to it. The direction of $V^a$ can be used to globally define a future
(or a past) direction for time.

Conversely, let $g_{ab}$ be a time-oriented
Lorentz metric, and at any point $p\in \cal M$ consider the set $\{ U^a\}$
of unit future-directed timelike vectors. A unique member of this
set will minimize $h_{ab}U^aU^b$. For, suppose that there are two
future-directed unit timelike vectors, $V^a_1$ and $V^a_2$, such that
$h_{ab}V^a_1V^b_1=K^2=h_{ab}V^a_2V^b_2$. Let $M=h_{ab}V^a_1V^b_2$ and
$L^2=-g_{ab}V^a_1V^b_2$. Then, $K^2>M$ and $L^2>1$.
It follows from this that any unit timelike vector of the
type $\hat V^a=\alpha V^a_1 + \beta V^a_2$, where $\alpha, \beta > 0$,
obeys $h_{ab}\hat V^a\hat V^b < K^2$. Thus the vector that minimizes
$h_{ab}U^aU^b$ must be unique. This gives the vector field $V^a$.

Both constructions ($g_{ab}$ and $V^a$) depend
on an entirely arbitrary choice of positive-definite metric~-- different
choices for $h_{ab}$ will give different Lorentz metrics (and different
vector fields).
The constructions are standard and I go into them only
because the existence of the vector field~$V^a$ is used repeatedly
as a tool in this paper: either to show that some manifold admits a
Lorentz metric, or, given a metric, to explore its properties.

If $\cal M$ has a boundary $\partial{\cal M}$,
some components of which are spacelike,
then $V^a$ cannot be tangent to these components.
Further, if~$\partial {\cal M}$ has a timelike component~$\cal T$,
then~$V^a$ can be deformed so as to be tangent to this component:
Let~$T^a$ be the timelike vector field tangent
to~$\cal T$ (chosen to be future-pointing)
and let $S^a_i$ ($i=1,\ldots,n-1$) be a set of mutually orthogonal
spacelike vectors, also orthogonal to~$T^a$. Choose these vectors
so that~$S^a_1$
is orthogonal to~$\cal T$ and the other~$S^a_i$ are tangent to it.
At each point $p\in\cal T$ let $\alpha_p$ be a small portion of the
spacelike geodesic with initial tangent~$S^a_1$.
Let~$s$ be a parameter on these curves, chosen to be zero at $\cal T$
(and to vary smoothly near $\cal T$).
The curves~$\alpha_p$ will not intersect each other close to~$\cal T$,
i.e, in some parameter range $[0,\,\epsilon_p)$.
Then $V^a$ may be modified close to $\cal T$ by replacing it with
$\hat V^a = (s/\epsilon_p)V^a + (1-s/\epsilon_p)T^a$ along the curves
$\alpha_p$. This, along with a suitable smoothing at~$\epsilon_p$,
is the required deformation of~$V^a$.

I will assume for the rest of this paper that $V^a$ is chosen to be
tangent to~$\cal T$ (at~$\cal T$).
This means that the integral curves of~$V^a$
must either lie in~$\cal T$ throughout or not intersect it all.

\SubSection Causal functions

We will also need to use some of the causal functions of
global general relativity~[\HE]. A $C^1$~curve is a curve $x^\mu(\tau)$
such that the components $dx^\mu/d\tau$ of the tangent to the curve exist
and are continuous.
A {\it timelike curve\/}
is defined to be a non-degenerate (i.e., not just a single point)
$C^1$ curve with a tangent that is everywhere timelike.
A {\it null curve\/} is similarly defined, except that degenerate curves
are
permitted here. A set is called {\it achronal\/} if no two points in it
can be connected by a timelike curve.
Timelike and null curves
are called {\it future-directed\/} if their tangents point somewhere
(and therefore, since they are $C^1$, everywhere) in the future time
direction. {\it Past-directed\/} curves are defined similarly.
Given a point $p\in \cal M$,
the {\it chronological future\/} of~$p$ is defined by
$$
I^+(p) =
\{ q\mid\exists \hbox{\sl \ a future-directed timelike curve from
$p$ to $q$}\}.
$$
The {\it chronological past\/} of~$p$, $I^-(p)$, is similarly
defined. The sets $I^{\pm}(p)$ are open sets. (There is a slightly
subtle point here: the ``openness'' of sets in the interpolating
manifold~$\cal M$ is with respect to the topology induced on it by the
full spacetime manifold in which~$\cal M$ lies. But
the interpolating
spacetimes that we are considering typically have boundaries;
in such cases an open set around a boundary point of~$\cal M$ will
itself, roughly speaking, ``have a partial boundary'' at its intersection
with~$\partial {\cal M}$.)
For ${\cal A}\subset\cal M$, the chronological
futures and pasts are defined by:
$$
I^\pm ({\cal A}) = \bigcup _{p\in{\cal A}}I^\pm (p).
$$
Clearly, these sets can be defined only in a time-oriented
spacetime. But an analogous set may be defined in general:
$$
I(p)=\{q\mid\exists \hbox{\sl \ a timelike curve between $p$ and $q$}\}.
$$
In a time-oriented spacetime we will have
$I(p)=I^+(p)\cup I^-(p)$.

\StartFigure{324}{324}
         {/coord {translate newpath 0 -72 moveto 0 72 lineto
                 2 69 lineto 0 72 moveto -2 69 lineto stroke
                 newpath -72 0 moveto 72 0 lineto
                 69 2 lineto 72 0 moveto 69 -2 lineto stroke} def
         newpath
         /point {1.5 0 360 arc gsave 0 setgray fill grestore} def
         newpath 102 272 moveto 142 312 lineto
         120 340 82 300 57 317 curveto closepath
         gsave .9 setgray fill grestore
         newpath 102 272 moveto 142 232 lineto
         142 190 lineto
         122 180 20 200 20 190 curveto closepath
         gsave .9 setgray fill grestore
         newpath 102 272 point
         newpath 20 190 moveto 142 312 lineto stroke
         newpath 142 232 moveto 57 317 lineto stroke
         102 272 point
         newpath
         180 272 moveto
         240 300 264 250 324 280 curveto
         gsave 1.5 setlinewidth stroke grestore
         newpath
         3 0 moveto
         72 67 72 67 141 0 curveto
         gsave 1.5 setlinewidth stroke grestore
         newpath
         0 0 moveto
         72 72 lineto 144 0 lineto stroke
         newpath
         92 82 moveto
         92 62 107 52 142 12 curveto
         gsave [5 5] 0 setdash stroke grestore
         newpath
         92 82 point
         newpath
         222 0 moveto 222 144 lineto 180 144 lineto 180 0 lineto closepath
         gsave .6 setgray fill grestore
         newpath
         282 0 moveto 282 144 lineto 324 144 lineto 324 0 lineto closepath
         gsave .6 setgray fill grestore
         newpath
         222 92 moveto
         256 126 lineto 282 102 lineto stroke
         newpath
         222 92 moveto
         242 97 262 87 282 102 curveto
         gsave 1.5 setlinewidth stroke grestore
         .5 setlinewidth
         gsave 72 72 coord grestore
         gsave 72 252 coord grestore
         gsave 252 72 coord grestore
         gsave 252 252 coord grestore
         }
\label(57,180){(a)}
\label(57,-15){(c)}
\label(237,180){(b)}
\label(237,-15){(d)}
\label(109,271){$p$} 
\label(90,302){$I^+(p)$}
\label(90,222){$I^-(p)$}
\label(272,261){$\cal S$} 
\label(217,300){$D^+({\cal S})$}
\label(217,225){$D^-({\cal S})$}
\label(20,62){$H^+({\cal S})\>\rightarrow$} 
\label(118.5,6){$\cal S$}
\label(40.5,34.5){$\matrix{\Big\uparrow\cr D^+({\cal S})\cr}$}
\label(95,84){$p$}
\label(137,25){$\mu$}
\label(259,85){$\cal S$} 
\label(223,125){$H^+({\cal S})$}
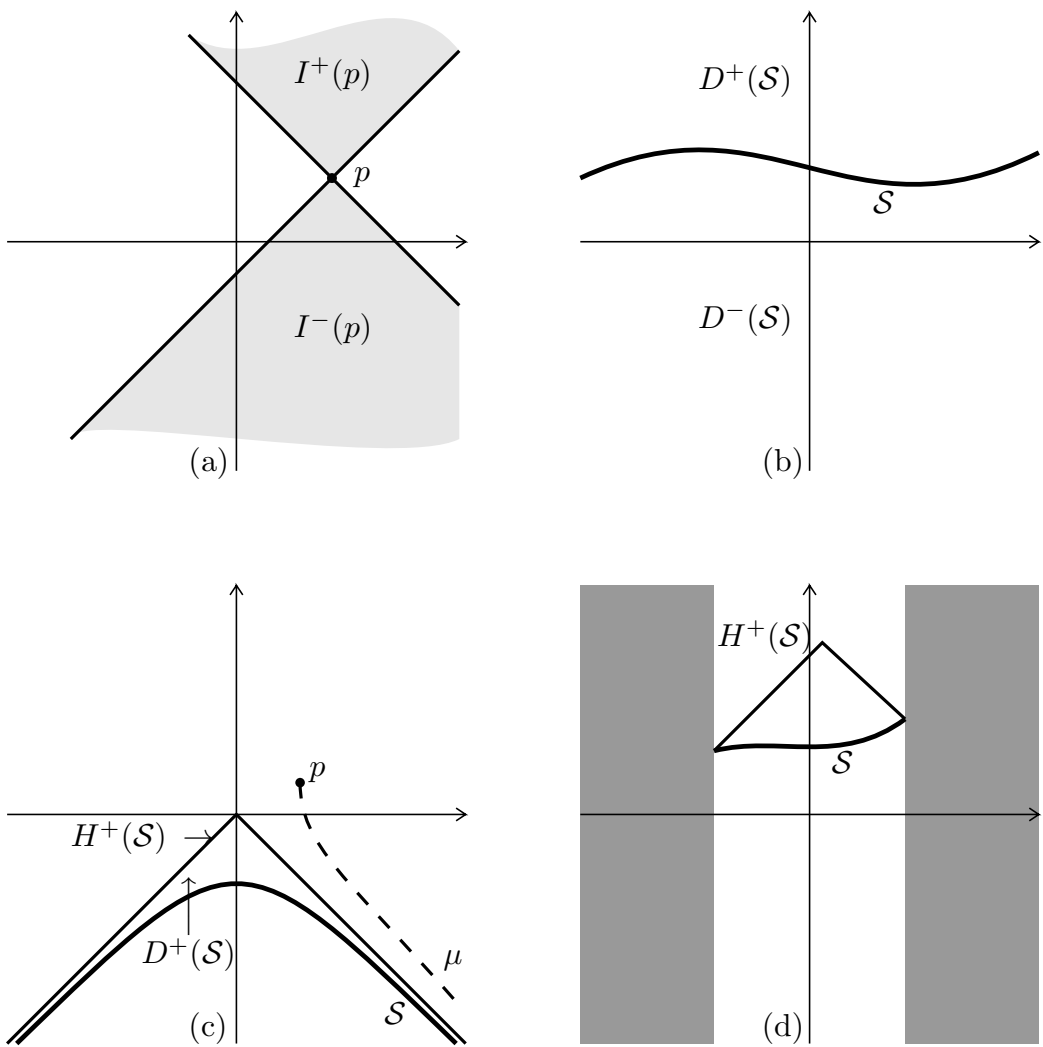
\caption{Some of the causal sets that are used in global
general relativity. All figures are based on 2-dimensional Minkowski
spacetime (with time pointing `upward').
Figure~(a) shows the future and past of a point~$p$. Figure~(b)
shows the future and past domains of dependance of a spacelike
surface~$\cal S$. Figure~(c) shows the future domain of dependance of
another spacelike surface~$\cal S$; here there is a Cauchy horizon because
there are points~$p$ to the future of~$\cal S$ from which there are
past-directed timelike curves
(such as the curve~$\mu$) that do not intersect~$\cal S$.
This Cauchy horizon exists because~$\cal S$ is badly chosen.
But in~(d), if the shaded regions are deleted,
then the existence of Cauchy horizons is intrinsic to the
truncated spacetime. Though this is a contrived example, pretty much
the same behavior occurs, for instance, in anti-de Sitter spacetime.}
\EndFigure

Another useful causal function is the {\it future domain of dependance\/} of
a connected achronal spacelike hypersurface~$\cal S$:
$$
\eqalign{
D^+({\cal S})&= \{ q\mid \hbox{\sl every past-directed timelike or null curve
from $q$}\cr
&\hphantom{=\{ q\mid\ }\hbox{\sl \ eventually meets $\cal S$ when extended
into the past}\}\cr}.
$$
The idea that this definition tries to capture is that the events that occur
in $D^+({\cal S})$ are determined entirely by initial data on $\cal S$.
The {\it past domain of dependance}, $D^-({\cal S})$, is defined similarly.
Now, clearly
$D^\pm ({\cal S})\subset I^\pm ({\cal S}$). In many cases it happens that
$\cal S$ is such (either because it is
badly chosen, or because of some intrinsic property of the spacetime) that
there are points to its future (i.e., in $I^+(\cal S$)) that do not
lie in $D^+({\cal S})$. Then $D^+({\cal S})$ has a future boundary, called
the {\it future Cauchy horizon}, defined as
$$
H^+({\cal S}) = \overline{D^+({\cal S})} - I^-(D^+({\cal S})).
$$
The {\it past Cauchy horizon}, $H^-(\cal S)$, is similarly defined.
Examples of most of these sets are shown in
\Fig. Various of their properties that are needed will be stated as the
occasion arises.

\Section The Kinematics of Topology Change I:\\
         Examples and Restrictions

Given two $(n-1)$-dimensional manifolds ${\cal S}_1$ and ${\cal S}_2$
(possibly of different topologies), under what conditions does there exist
an interpolating spacetime $\cal M$ between them? If $\cal M$ exists, what
properties must it have?

\StartFigure{180}{216}
           {newpath
           gsave
           /manifold {translate newpath -25 0 moveto
                      -15 60 20 55 30 0 curveto
                      40 -55 15 -30 5 -30 curveto
                      -20 -30 -35 -60 -25 0 curveto
                      stroke grestore} def
           gsave 18 170 manifold
           gsave 152 170 manifold
           gsave 37 36 manifold
           gsave 143 36 manifold
           grestore
           newpath
           18 170 14 0 360 arc gsave .6 setgray fill grestore
           newpath
           152 170 14 0 360 arc gsave .6 setgray fill grestore
           newpath
           46 36 14 90 270 arc
           94 30 112 30 134 22 curveto
           134 36 14 270 90 arc
           112 32 94 32 46 50 curveto
           gsave 1 setgray fill grestore
           newpath
           gsave [2 2] 0 setdash
           46 36 14 90 270 arc stroke grestore
           newpath
           46 22 moveto
           94 30 112 30 134 22 curveto stroke
           newpath
           gsave [2 2] 0 setdash
           134 36 14 270 90 arc stroke grestore
           newpath
           134 50 moveto
           112 32 94 32 46 50 curveto stroke
           .5 setlinewidth
           newpath
           90 130 moveto
           90 100 lineto
           93 105 lineto
           90 100 moveto
           87 105 lineto stroke
           }
\label(-26,180){${\cal M}_1$}
\label(187,180){${\cal M}_2$}
\label(71,80){${\cal M}_1\# {\cal M}_2$}
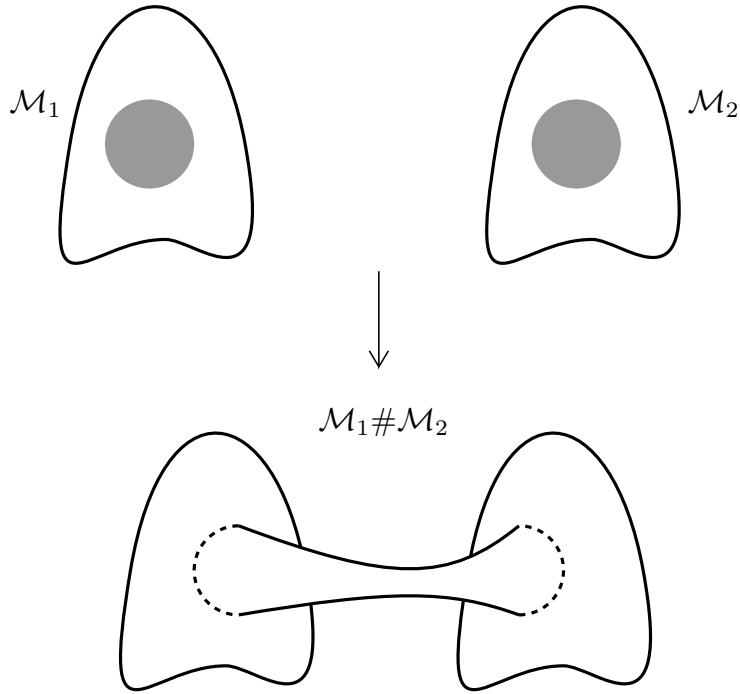
\caption{The connected sum of two manifolds ${\cal M}_1$ and ${\cal M}_2$.
         An open disk~-- represented by a shaded region above~-- is removed
         from each of the manifolds and the edges of the two disks are
         identified.}
\EndFigure

\SubSection Connected sums

To discuss the first of these questions, it is useful to have on hand the
following technique for constructing new manifolds out of old. Given any
two $n$-dimensional manifolds ${\cal M}_1$ and ${\cal M}_2$
(with or without boundary), their {\it connected sum\/} is formed by deleting
an
open $n$-disc from the interior of each of ${\cal M}_1$ and ${\cal M}_2$ and
identifying them along the resulting
boundaries~[\SorkOne\,
\StoreRef{\Yodz}
\Refe{P. Yodzis, \Jou{Comm. in Math. Phys.}, \Vol{26}, 39 (1972);
\Jou{Gen. Rel. and Grav.}, \Vol{4}, 299 (1973).}.
The process is illustrated in~\Fig.

Using this construction it is possible to show that if no restrictions
are placed on the interpolating spacetime,
then it always exists. The existence, first,
of an interpolating manifold is easily seen: given $(n-1)$-dimensional
manifolds ${\cal S}_1$ and ${\cal S}_2$, let
${\cal M}_i = {\cal S}_i\times [0,\,\infty )$;
then ${\cal M}'={\cal M}_1 \# {\cal M}_2$ is the required manifold.
This can now be modified to admit a Lorentz metric. Observe
first that each of ${\cal M}_1$ and ${\cal M}_2$
admits a natural vector field along
the $[0,\,\infty )$ lines. Pick the vector fields to point into ${\cal M}_1$ on
${\cal S}_1$ (i.e., in the direction of increasing $t\in [0,\,\infty )$ on
${\cal M}_1$) and out of ${\cal M}_2$ on ${\cal S}_2$
(i.e., in the direction of decreasing $t\in [0,\,\infty )$ on ${\cal M}_2$).
On ${\cal M}_1$, let ${\cal D}_1$ be the disc that
was removed for the purposes of constructing ${\cal M}'$. Cut out a shell
almost completely surrounding ${\cal D}_1$ and modify the
field near ${\cal D}_1$ so as to point into it (see \NFig). Do the same
with ${\cal M}_2$, but this time modify the vector field so that it points
out of ${\cal D}_2$. The new connected sum, $\cal M$, will have a vector
field on it from which a Lorentz metric may be constructed.
${\cal S}_1$~will be an initial surface and ${\cal S}_2$ a final surface
for this metric (if the direction of the
vector field is used to define the future direction of time).
\StartFigure{328}{164}
          {/arrow {newpath 0 0 moveto
                   0 20 lineto
                   3 15 lineto
                   0 20 lineto
                   -3 15 lineto gsave .5 setlinewidth stroke grestore} def
           /spacetime {
           newpath 0 0 moveto
           40 20 104 -10 144 0 curveto stroke
           gsave 8 15 translate arrow grestore
           gsave 40 15 translate arrow grestore
           gsave 72 15 translate arrow grestore
           gsave 104 15 translate arrow grestore
           gsave 136 15 translate arrow grestore
           gsave 22 50 translate arrow grestore
           gsave 122 50 translate arrow grestore
           gsave 8 85 translate arrow grestore
           gsave 136 85 translate arrow grestore
           gsave 22 120 translate arrow grestore
           gsave 56 120 translate arrow grestore
           gsave 88 120 translate arrow grestore
           gsave 122 120 translate arrow grestore
           newpath
           72 72 20 0 360 arc gsave .6 setgray fill grestore
           gsave
           2 setlinewidth
           newpath
           72 72 36 -60 240 arc stroke
           grestore
           gsave 106 72 translate 90 rotate 1 .6 scale arrow grestore
           gsave 38 72 translate -90 rotate 1 .6 scale arrow grestore
           gsave 72 106 translate -180 rotate 1 .6 scale arrow grestore
           gsave 72 72 translate 30 rotate 0 -34 translate
           1 .6 scale arrow grestore
           gsave 72 72 translate -30 rotate 0 -34 translate
           1 .6 scale arrow grestore } def
           spacetime
           /arrow {newpath 0 20 moveto
                   0 0 lineto
                   3 5 lineto
                   0 0 lineto
                   -3 5 lineto gsave .5 setlinewidth stroke grestore} def
           324 150 translate 180 rotate spacetime
        }
\label(-10,5){${\cal S}_1$}
\label(-15,80){${\cal M}_1$}
\label(324,157){${\cal S}_2$}
\label(334,80){${\cal M}_2$}
\label(86,88){${\cal D}_1$}
\label(227,57){${\cal D}_2$}
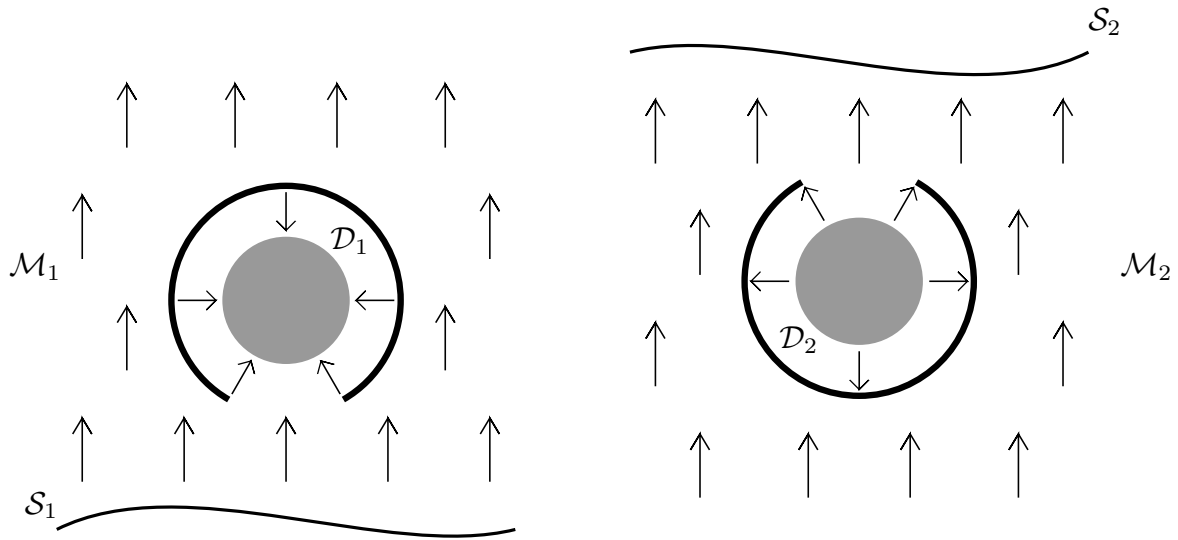
\caption{How the timelike vector field is set up on ${\cal M}_1$ and
${\cal M}_2$. The discs ${\cal D}_1$ and ${\cal D}_2$ (shown shaded),
and the almost-complete shells surrounding them (shown as thick arcs),
are deleted from the manifold. The edges of the two discs are identified
in the standard connected-sum construction illustrated earlier.}
\EndFigure

\SubSection Causal compactness

The construction discussed above is very arbitrary. Since the
interpolating spacetime~$\cal M$
is not constrained in any way, we could continue making
cuts and identifications in it as we choose. This is not a very
satisfactory state of affairs and it is important, therefore, to restrict
$\cal M$ in some reasonable manner. The sort of restriction that is desirable
would be one that does not allow points of $\cal M$ to have causal access
to holes or to ``regions at infinity.'' Such a restriction will have to be
made on the interpolating spacetime, not just on the underlying manifold.
Consider, for
example, the manifold $[0,\,1]\times R^3$. A Minkowski metric can be put
on it such that the two boundaries correspond to constant-$t$ surfaces
(where $t$~is the usual time co-ordinate). There are no holes in this
spacetime, and no point in it can be causally connected to a region at
infinity. But, the same manifold also admits an anti-de Sitter metric. A
time coordinate may be picked here such that the two boundaries
again correspond to constant-$t$ surfaces, but with points in $\cal M$
now able to receive signals from points at infinity
(ref.~[\HE], p.~131--134).
(Fig.~2d illustrates such a spacetime. Between any two constant-$t$
surfaces in it, however close they are to each other, there will be points
that can receive signals from the boundary with the deleted region.)
Now, this particular class of situations may
be handled by the freedom that we have left ourselves in the definition
of an interpolating spacetime: since the boundary~$\partial M$ may
contain a further component, the timelike tube $\cal T$, the region near
infinity can be put outside this tube. I will return
to this when I discuss open universes. But the general problem
of arbitrary cuts and holes in~$\cal M$ is more serious: it cannot
be handled, in general, by putting all the ``bad parts'' outside
timelike tubes and concentrating only on what happens inside.
Such problematic
situations can all be excluded, however, if we require that the
interpolating spacetime obey a condition
called {\it causal compactness}. This condition is meant to take away
the license to arbitrarily make cuts and holes in spacetime:

\Declare{Definition} A spacetime $\cal M$ is called causally compact
if for any $p\in \cal M$, $\overline{I(p)}$ is compact.

\StartFigure{210}{154}
        {  
           newpath
           0 10 moveto
           70 0 140 20 210 10 curveto
           210 144 lineto
           120 114 50 144 0 134 curveto closepath
           clip
           newpath
           0 -20 moveto
           100 80 lineto
           200 -20 lineto
           closepath gsave .9 setgray fill grestore
           newpath
           200 180 moveto
           100 80 lineto
           0 180 lineto
           closepath gsave .9 setgray fill grestore
           newpath
           0 -20 moveto
           200 180 lineto stroke
           newpath
           200 -20 moveto
           0 180 lineto stroke
           100 80 1.5 0 360 arc gsave 0 setgray fill grestore
           newpath
           0 10 moveto
           70 0 140 20 210 10 curveto stroke
           newpath
           0 134 moveto
           50 144 120 114 210 144 curveto stroke
         }
\label(107,80){$p$}
\label(0,16){${\cal S}_1$}
\label(0,141){${\cal S}_2$}
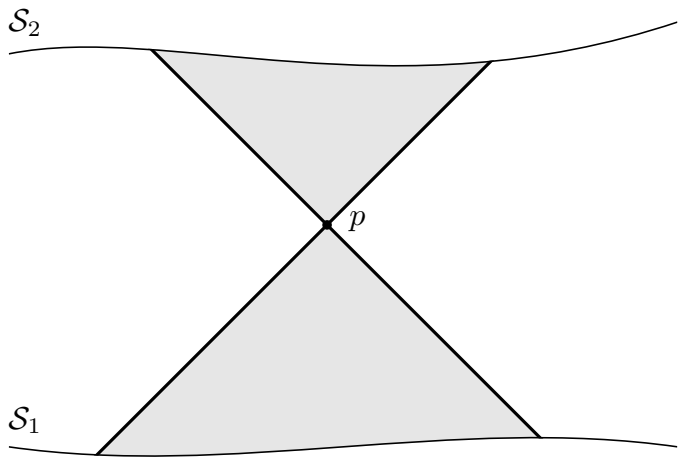
\caption{A causally compact spacetime. Points~$p$ between
         ${\cal S}_1$ and ${\cal S}_2$ cannot receive signals from, or send
         signals to, either regions at infinity or ``holes'' in the
         spacetime.}
\EndFigure

This definition has been phrased in such a way as to be applicable both to
spacetimes that are not time-orientable, as well as to ones that are.
If the spacetime is time-orientable we have $I(p)=I^+(p)\cup I^-(p)$.
Thus~$\overline{I^+(p)}$ and~$\overline{I^-(p)}$, being closed subsets of
the compact set~$\overline{I(p)}$, must each be compact.
The idea of causal compactness is illustrated in \Fig.

There are many classes of situations in which ${\cal S}_1$ and ${\cal S}_2$
can be connected by a causally compact interpolating spacetime. They
are studied systematically below.

\SubSection Closed universes

In a closed universe,
${\cal S}_1$ and ${\cal S}_2$ are (by definition)
both closed surfaces (i.e., compact without boundary) and it is reasonable
to require that $\cal M$ be compact as well. (Thus there is no timelike
component to the boundary of~$\cal M$ here; i.e., ${\cal T} = \emptyset$.)
In this case, $\cal M$ will trivially be causally compact with respect
to any Lorentz metric that it admits. To study such situations we can
draw on results from cobordism theory
\StoreRef{\MilnStash}
\Ref{J.W. Milnor and J.D. Stasheff, \Jou{Characteristic Classes},
Princeton University Press, Princeton (1974).}.
If a compact manifold
$\cal M$ interpolates between closed manifolds ${\cal S}_1$ and ${\cal S}_2$,
then $\cal M$ is called a {\it cobordism}, and ${\cal S}_1$ and ${\cal S}_2$
are called {\it cobordant}. The condition for ${\cal S}_1$ and ${\cal S}_2$
to be cobordant is that their Stiefel-Whitney numbers (these numbers
characterize some of the properties of a manifold) be the same~[\MilnStash].
If ${\cal S}_1$, ${\cal S}_2$ and $\cal M$ are required to be
oriented, then their Pontryagin numbers must be equal as well~[\MilnStash].
These requirements are satisfied in a wide variety of cases:
e.g., when $n = 2,3,4,7$ or~$8$, any two $(n-1)$-dimensional closed oriented
manifolds are cobordant through an oriented cobordism.
In fact, when $n=4$ (which, in the naive past, used to be considered the case
of greatest physical interest),
any two closed manifolds, oriented or not, are cobordant.

Given the existence of the cobordism $\cal M$, the next question
is, is it {\it Lorentzian\/}? i.e., can
an appropriate Lorentz metric be put on it?
Equivalently, we can ask under
what conditions it is possible to put a vector field on $\cal M$ that is
nowhere tangent to ${\cal S}_1$ and ${\cal S}_2$. The condition for
such a field to exist and to point outward everywhere (or inward
everywhere) on
$\partial \cal M$ is that $\chi ({\cal M}) = 0$, where $\chi (\cal M)$
is the Euler characteristic of $\cal M$
\StoreRef{\Milnor}
\Ref{J.W. Milnor, \Jou{Topology from the Differentiable Viewpoint},
Univ. Press of Virginia, Charlottesville (1965).}.
We are, however, more interested in the case when the field
points into $\cal M$ on ${\cal S}_1$ and out of $\cal M$ on ${\cal S}_2$.
The conditions for the existence of such a field have also been obtained
before, by Reinhart
\StoreRef{\Rein}
\Ref{B.L. Reinhart, \Jou{Topology}, \Vol{2}, 173 (1963).}
and by Sorkin~[\SorkTwo]:
{\sl
Let $n$ be the dimension of $\cal M$. If~$n$ is even, then $\cal M$ admits
such a field if $\chi ({\cal M}) = 0$; if~$n$ is odd, then $\cal M$
admits the field if $\chi ({\cal S}_1) = \chi ({\cal S}_2)$.}

Observe that these conditions include the condition for the vector field to
point outward everywhere (or inward everywhere) on $\partial \cal M$.
For, that case can be regarded as representing the topological transition
$\emptyset\rightarrow\partial{\cal M}(={\cal S}_1\cup{\cal S}_2)$.
Then, both when~$n$ is even as well as when it is odd, the condition
$\chi ({\cal M})=0$ can be recovered (using the result that
$2\chi ({\cal M})=\chi (\partial {\cal M})$ if $\cal M$ is odd-dimensional).

The condition when $n$~is odd serves as a selection rule, which I will
refer to as the Reinhart-Sorkin selection rule (or the RS rule, for short),
and it forbids a
number of topological transitions. To see this,
consider the first few cases. (In the discussion below, $S^n$ and $T^n$ will
stand for the $n$-dimensional sphere and torus respectively.)

\Bull{$n=1$}:
This case is not very interesting, for
there are topologically only four 1-manifolds~[\Milnor]:
$S^1$ (i.e., the circle); $R^1$; $[0,\,\infty)$; and $[0,\,1]$.

\Bull{$n=3$}:
The RS rule forbids all Lorentzian
topological transitions between oriented closed 2-manifolds (except, of
course, the identity transition). This is so, because any such manifold may
be obtained by adding a certain number of handles to~$S^2$. The number of
handles is called the {\it genus}, g, and it completely characterizes the
topology of the (oriented) manifold.
($S^2$ has g = 0; $T^2$ has g = 1; etc.)
The Euler characteristic is related to the genus by $\chi = 2-2$g
and is, therefore, different for oriented closed 2-manifolds of different
topologies.

\Bull{$n=5$}:
Although some topological transitions may be possible here, Sorkin has
shown~[\SorkTwo] that the RS~rule does not allow, for example, monopole
pair creation in (5-dimensional) Kaluza-Klein theory.

\Bull{$n=7$}:
Here is one example of topology change: $S^6$ and
$(S^4\times S^2)\# T^6$ are cobordant and
both have $\chi = 2$. But the first of these is simply connected
and the second is not, so the transition between the two
indeed represents topology change.

\smallskip
In the even-dimensional case it is the interpolating spacetime that is
restricted, not the initial and final surfaces. This suggests a further
question: if a given cobordism
does not have $\chi = 0$, can it be modified in some way so that $\chi$
now vanishes? Such a modification can be carried out by imitating a procedure
due to Misner, and
used by Geroch in 4~dimensions~[\Gero\,\GeroPhD]. The procedure is based
on the following standard even-dimensional result:
$$
\hbox{If\ } {\cal N} = {\cal M}\# {\cal V}, \hbox{\ then\ }
\chi ({\cal N}) = \chi ({\cal M}) + \chi ({\cal V}) - 2.
$$
Now, if ${\cal V} = T^n$, then $\chi ({\cal V}) = 0$;
and if ${\cal V} = S^2\times S^{n-2}$ (for $n>2$), then
$\chi ({\cal V}) = 4$. Thus, if $\chi (\cal M)$
is even to start with (in dimensions $> 2$), then
by repeated application of \# with the appropriate $\cal V$ it can be reduced
to zero. If $\chi (\cal M)$ is odd initially, and if $n = 4k$, we can use
${\cal V} = CP^2\times CP^2\times ...$ ($k$ factors). Here $CP^2$ is complex
projective 2-space, which is a real 4-dimensional closed manifold with
$\chi = 3$. So $\chi ({\cal V}) = 3^k$; this $\cal V$ can
be used to first make a manifold
of even $\chi$, to which the above procedure can then be applied. If $n=4k+2$,
then there are no orientable closed manifolds of odd $\chi$ for us to use.
But, if orientability is not a concern, then $RP^n$ with $\chi = 1$ may
be used in place of $CP^2$. And, finally, in the two dimensional case it is
possible to enumerate all the possibilities explicitly~[\SorkTwo].
They are listed separately below.

To summarize these results, suppose that
${\cal S}_1$ and ${\cal S}_2$ are cobordant and
of dimension~$(n-1)$. Then, if~$n$ is odd, there is a spacetime that
interpolates between them if and only if
$\chi ({\cal S}_1) = \chi ({\cal S}_2)$~[\SorkTwo],
and examples exist of such Lorentzian cobordisms. If~$n=2$, then again
examples of topology change can be constructed, as shown below.
And, if~$n$ is even and is greater than~$2$,
then there is always an interpolating spacetime. (This result
was found previously, with slightly different methods~[\Rein].)
Thus, kinematically, though there
are some constraints, a large variety of topology-changing closed universes
exist. In particular, as Misner has pointed out~(see [\Gero]),
since any two closed 3-manifolds are cobordant, there is at this stage
no barrier whatsoever to four-dimensional topology change.

(It is worth noting here that Gibbons and Hawking~[\GibHawk] have recently
found additional selection rules that further constrain the possible
topological transitions that can occur.)

\SubSection Two-dimensional topology change

The only Lorentzian cobordisms in two dimensions
are the torus and the Klein bottle (both with no
boundaries), the cylinder (with two $S^1$~boundaries), and the M\"obius
strip (with one $S^1$~boundary)~[\SorkTwo].
The first two may be regarded as examples
of a $\emptyset\leftrightarrow\emptyset$ transition. On the cylinder, the
most obvious choice of metric makes it an example of the identity
transition, $S^1\leftrightarrow S^1$ (\NFig a);
but a metric may also be chosen on it (\NFig b)
so that it represents the transition $S^1\cup S^1\leftrightarrow\emptyset$.
And on the
M\"obius strip, one choice of metric (\NFig c) makes it an example of an
$S^1\leftrightarrow\emptyset$ transition~[\SorkTwo]; another choice
(non-time-orientable) is discussed in section~VII. The spacetimes
of \NFig b and \NFig c are explicit examples of Lorentzian topology change.
\smallskip
\StartFigure{240}{170}
          {/arrow {newpath 0 0 moveto
                   0 10 lineto
                   2 7 lineto
                   0 10 lineto
                   -2 7 lineto gsave .7 setlinewidth stroke grestore} def
           /botcirc {gsave
           newpath
           1 .25 scale
           1.6 setlinewidth
           20 0 20 180 360 arc stroke
           [1 1] 0 setdash
           newpath
           20 0 20 0 180 arc stroke
           grestore } def
           newpath
           gsave
           1 .25 scale
           1.6 setlinewidth
           20 640 20 0 360 arc stroke
           grestore
           newpath 0 160 moveto
           0 20 lineto stroke
           newpath 40 160 moveto
           40 20 lineto stroke
           gsave 0 20 translate botcirc grestore
           20 40 140 {/yc exch def
                      gsave 10 yc translate arrow grestore} for
           20 40 140 {/yc exch def
                      gsave 30 yc translate arrow grestore} for
           40 0 translate
           gsave 80 90 translate botcirc grestore
           gsave 160 90 translate botcirc grestore
           newpath 120 90 moveto
           110 162 170 162 160 90 curveto stroke
           newpath 80 90 moveto
           80 202 200 202 200 90 curveto stroke
           gsave 90 90 translate arrow grestore
           gsave 110 90 translate arrow grestore
           gsave 170 90 translate arrow grestore
           gsave 190 90 translate arrow grestore
           gsave 139 165 translate -180 rotate arrow grestore
           gsave 100 130 translate -60 rotate arrow grestore
           gsave 180 130 translate 60 rotate arrow grestore
           gsave
           newpath
           1 .5 scale
           1.3 setlinewidth
           140 100 40 -90 180 arc stroke
           newpath
           140 40 40 -120 0 arc stroke
           newpath
           140 40 40 55 153 arc stroke
           newpath
           140 40 40 170 200 arc stroke
           grestore
           newpath
           100 50 moveto
           100 35 100 8 120 3 curveto stroke
           newpath
           108 15 moveto
           120 30 130 30 140 30 curveto stroke
           /VField {
           gsave 140 2 translate -45 rotate arrow grestore
           gsave 140 28 translate -135 rotate arrow grestore
           gsave 150 16 translate -75 rotate arrow grestore } def
           VField
           gsave 280 70 translate 180 rotate VField grestore
           newpath
           180 20 moveto 180 50 lineto [3 3] 0 setdash stroke
         }
\label(-17,20){(a)}
\label(100,90){(b)}
\label(120,0){(c)}
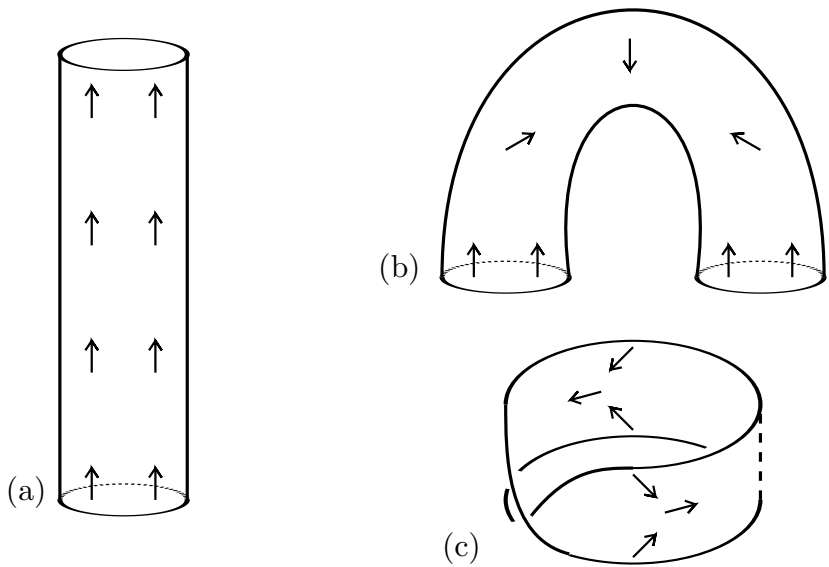
\caption{Examples of 2-dimensional Lorentzian cobordisms.
The manifolds in~(a) and~(b) are cylinders, and the one in~(c) a
M\" obius strip. (The spacetime of fig.~(c) is due to Sorkin~[\SorkTwo].)
The arrows show the vector field
$V^a$, from which the Lorentz metric may be constructed. The field
$V^a$ may be thought of as representing the future direction of time.
The pictures in~(b) and~(c) are examples of topology
change. They represent, respectively, the transitions $S^1 \cup S^1
\rightarrow \emptyset$, which may be thought of as universe
pair-annihilation, and $S^1 \rightarrow \emptyset$, which may be thought
of as universe self-annihilation. Reversing~$V^a$
gives the reverse transitions: universe pair-creation and universe
self-generation.}
\EndFigure

\SubSection Open universes

Next, consider the case when ${\cal S}_1$ and ${\cal S}_2$ are non-compact.
It seems reasonable to expect that many of these situations will also be
causally compact, with $\cal T=\emptyset$, as they stand.
For example, Minkowski spacetime is, and so are the open
Friedmann cosmologies (if, in both cases, the boundaries ${\cal S}_1$ and
${\cal S}_2$ correspond to constant-$t$ surfaces, where~$t$ is the
usual time coordinate). But the freedom that comes from the
extra component, $\cal T$, of the boundary of~$\cal M$ allows us now to
embrace even wider classes of situations.

One such situation of interest is the asymptotically flat spatial geometry
of an isolated system. We expect to be able to compactify this situation by
adding a point at infinity, or by imposing periodic boundary conditions
(``putting it in a box''), and thus we might expect
results similar to the compact
case. For the 4-dimensional case Geroch has defined an {\it externally
Euclidean\/} 3-manifold $\cal S$ to be one that contains a
compact set $\cal C$ such that ${\cal S}-{\cal C}$ is diffeomorphic
(see section~VIII for a definition of the term) to
$R\times S^2$. It is reasonable to assume that an isolated system can be
represented thus. Then, given two externally Euclidean 3-manifolds
${\cal S}_1$ and ${\cal S}_2$, a spacetime $\cal M$ that
interpolates between them is called {\it externally Lorentzian\/} if there is a
compact set $\cal K$ such that ${\cal M}-\cal K$
is diffeomorphic to $S^2\times R \times [1,\,2]$ where
${\cal S}_t = S^2\times R\times\{ t\} $ is spacelike for each $t\in [1,\,2]$,
and $\gamma_p = \{ p\} \times [1,\,2]$ is timelike and future-directed for each
$p\in S^2\times R$. The idea behind this definition is that the
topology change, if any, occurs within the compact set $\cal K$. Then, in
4 dimensions, in exact analogy with the compact case, any two externally
Euclidean spaces can be connected by an interpolating externally
Lorentzian spacetime~[\GeroPhD].

Though this definition may be extended in a straightforward way to
arbitrary numbers of dimensions, it is a little restrictive.
The structure of an asymptotically flat space ``close to infinity''
in higher-dimensional theories would not, for instance, necessarily be
$R\times S^{n-2}$. We might, in many cases, expect instead
something like $R\times S^2\times \cal N$ where $\cal N$ is the
compact ``internal manifold.'' In fact, examples of both kinds of behaviour
at infinity are known~[\BGR]. It is also possible that we might have more
complicated structures, not expressible as product manifolds globally.
Another class of situations not necessarily covered above is
the multi-geon class
involving infinitely many topological kinks.

It is reasonable to suppose that all of these scenarios belong to a
class of spacetimes that may be called {\it externally simple\/}:

\Declare{Definition} A spacetime $\hat{\cal M}$ that interpolates between
two spacelike surfaces $\hat{\cal S}_1$ and $\hat{\cal S}_2$ is called
externally simple if
\item{a.} $\hat{\cal M}$ contains a (possibly empty) causally compact region
whose initial and final surfaces, ${\cal S}_1$ and ${\cal S}_2$,
are subsets of $\hat{\cal S}_1$ and $\hat{\cal S}_2$, respectively, and
\item{b.} $\hat{\cal S}_1 - {\cal S}_1$ is diffeomorphic to
$\hat{\cal S}_2 - {\cal S}_2$.

The idea here is that nothing topologically interesting happens outside the
causally compact part of an externally simple spacetime.

The results in the rest of this paper all deal with
causally compact interpolating spacetimes. They are meant to apply to
situations where the timelike component~$\cal T$ of~$\partial{\cal M}$
is empty, or, if non-empty, where~$\cal M$ is a causally compact region
of a larger, externally simple spacetime~$\hat{\cal M}$.

\Section The Kinematics of Topology Change II:\\
         Causality Violations

We have seen so far that there appear to be few kinematical obstructions
to topology change in general relativity. The question that next arises
is, what are the properties of topology-changing spacetimes?
It turns out that these
properties are not pleasant. The first sign of this
came with Geroch's discovery that causality violations
are necessarily associated with topology change in the closed universe and
externally Lorentzian cases~[\Gero\,\GeroPhD].
By ``causality violation'' it is meant that
there is at least one closed timelike curve. This will happen, for instance,
if there are points $q_1$ and $q_2$ such that $q_1\in I^-(q_2)$ and
$q_2\in I^-(q_1)$.

The same result may be proved for causally compact
interpolating spacetimes in general. The proof that I give below is
slightly different from Geroch's.

\Theorem
Let $\cal M$ be a (time-oriented) causally compact spacetime that
interpolates between an initial surface ${\cal S}_1$ and a final surface
${\cal S}_2$. Suppose that $\cal M$ has no closed timelike curves. Then,
$\cal M$ is diffeomorphic to ${\cal S}_1\times [0,\,1]$ (and, in particular,
${\cal S}_1$ is diffeomorphic to~${\cal S}_2$).

(A variant of this result, but now not assuming time-orientability,
is discussed in section~VII.)

\Proof
Because $\cal M$ is time-oriented, $I(p)=I^+(p)\cup I^-(p)$
for any point $p\in {\cal M}$.
So, by causal compactness, each of
$\overline{I^+(p)}$ and $\overline{I^-(p)}$ must separately be compact.
Let $V^a$ be a smooth, future-directed timelike vector field on $\cal M$
(chosen to be tangent to the timelike component of $\partial{\cal M}$,
$\cal T$, if such a timelike component exists).
If $p\in {\cal M}$, let $V^a_p$ denote the element of this vector
field at~$p$. The field $V^a$ will give rise to a set of unique,
smooth integral curves (i.e., curves to which $V^a$ is tangent~--
more precisely, solutions of $(dx^a/dv)=V^a$) on~$\cal M$
\StoreRef{\CodLev}
\Ref{E.A. Coddington and N. Levinson, \Jou{Theory of Ordinary Differential
Equations}, McGraw Hill, New York (1955).}.
The main thrust of the proof is to show that in the absence
of closed timelike curves,
each of the integral curves of~$V^a$ must have a past
endpoint on~${\cal S}_1$ and a future endpoint on~${\cal S}_2$.

The curves that lie in~$\cal T$ take on such endpoints by definition.
Suppose that one of the other curves $\lambda$
(i.e., $\lambda\cap{\cal T}=\emptyset$)
does not have (say) a future endpoint on~${\cal S}_2$.
A contradiction ensues, as shown below:

First, observe that $\lambda$ cannot have a future endpoint anywhere
else on~$\cal M$ (since $V^a$ is defined everywhere).
Let $b\in\lambda$ and let~$\mu$ be the portion of~$\lambda$ to the
future of~$b$. Then $I^+(\mu)\subset I^+(b)$. Now
$\overline{I^+(b)}$ is compact
by assumption; so, since $\overline{\mu} \subset
\overline{I^+(\mu)} \subset \overline{I^+(b)}$, $\overline{\mu}$
must also be compact.
Consider the set ${\cal N} = \bigcup_{p\in \mu} I^-(p)$.
Clearly $\mu\subset\cal N$. I now show that, further,
$\overline{\mu}\subset\cal N$.

Suppose that there is a sequence
of points $q_i$ on $\mu$ that converges to a point~$q$ not on $\mu$.
Then the vectors
$V^a _{q_i}$ will converge to the vector $V^a _q$. Let $\rho$ be the
integral curve with initial tangent $V^a _q$ at $q$ and let $q'$ lie to the
future of $q$ on $\rho$. Since $I^+(q)$ is open, all points sufficiently close
to $q'$ will
also lie in $I^+(q)$. Some of these points will lie on integral curves of
$V^a$ through the $q_i$ close to $q$, i.e., they will lie on $\mu$.
Thus, there will
be some $p\in\mu$ such that $q\in I^-(p)$, i.e., $q\in \cal N$.
Therefore, $\cal N$ provides an open covering of the compact set
$\overline{\mu}$. Let
$\cal N$$' = \bigcup_{i=1}^{n} I^-(p_i)$ be a finite subcovering and suppose
that $p_k$ is the futuremost of the points $p_i$ on $\mu$. But,
$p_k\in\mu\subset \cal N$$'$, i.e., $p_k\in I^-(p_r)$ for some $r$.
But we also have $p_r\in I^-(p_k)$. This gives a closed timelike
curve.
\StartFigure{144}{100}
          {
           newpath
           0 10 moveto
           30 0 114 20 144 0 curveto stroke
           newpath
           0 80 moveto
           30 90 114 70 144 90 curveto stroke
           newpath 40 40 moveto
           40 80 60 80 60 40 curveto
           60 10 75 10 75 40 curveto
           75 80 85 80 85 40 curveto
           85 10 91 10 91 40 curveto
           91 80 96 80 96 40 curveto
           96 10 99 10 99 40 curveto
           99 80 101 80 101 40 curveto
           101 10 103 10 103 40 curveto
           stroke
           newpath
           106 40 1 0 360 arc fill
           newpath
           40 25 moveto 40 40 lineto stroke
           newpath
           .8 setlinewidth
           40 10 moveto
           40 22 lineto
           37 18 lineto 40 22 lineto 43 18 lineto stroke
        }
\label(0,17){${\cal S}_1$}
\label(0,88){${\cal S}_2$}
\label(24,13){$V^a$}
\label(31,50){$\lambda$}
\label(108,40){$q$}
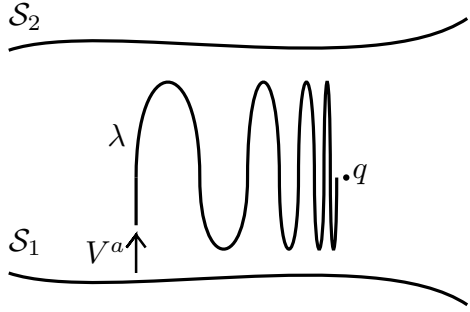
\caption{A sketch of how the proof of theorem~1 runs. (Though the
surfaces ${\cal S}_1$ and ${\cal S}_2$ are spacelike, time does not
point upward everywhere in the region between the surfaces.)
If ${\cal S}_1$
and ${\cal S}_2$ are not diffeomorphic, at least one integral curve
of~$V^a$, $\lambda$, must wind around trapped in the interpolating spacetime.
By causal compactness, the curve accumulates at some point~$q$, and this
accumulation leads to a closed timelike curve.}
\EndFigure

So, every integral curve of the field $V^a$ takes on an endpoint on
${\cal S}_1$ and on ${\cal S}_2$. We may choose a parameter on each
curve that has value~0 at ${\cal S}_1$ and~1 at ${\cal S}_2$.
Since~$V^a$ is smooth, this means that $\cal M$ is
diffeomorphic to ${\cal S}_1\times [0,\,1]$.
\EndProof

This theorem also makes it possible to make precise the
result that was mentioned in the introduction: a spacetime cannot
give topology change if it is without
holes (i.e., it is causally compact) and can be foliated by
spacelike hypersurfaces.

\Declare{Corollary}
Let $\cal M$ be a (time-oriented) causally compact spacetime that
interpolates between an initial surface ${\cal S}_1$ and a final surface
${\cal S}_2$. Suppose that $\cal M$ can be foliated by
spacelike surfaces, with~${\cal S}_1$ as the first surface in the
foliation and~${\cal S}_2$ as the last. Then,
$\cal M$ is diffeomorphic to ${\cal S}_1\times [0,\,1]$ (and, in particular,
${\cal S}_1$ is diffeomorphic to~${\cal S}_2$).

\Proof
Since~$\cal M$ can be foliated in the manner described above, it may be
expressed as
$$
\bigcup_{t\in [1,\,2]} {\cal S}_t,
$$
where each ${\cal S}_t$ is a spacelike hypersurface and
${\cal S}_{t_1}\cap{\cal S}_{t_2}=\emptyset$ for $t_1\ne t_2$.
Then ${\cal M}$ can contain no closed timelike curves
(since~$t$ must strictly increase along every future-directed timelike
curve), and the result follows from \PrevThm.
\EndProof

Since, by assumption, $\cal M$ is connected, it follows from the conditions
of~\PrevThm\ that ${\cal S}_1$ and ${\cal S}_2$ must also be
connected. A similar restriction also follows from slightly weaker
assumptions (the proof is fashioned after one originally given by
Geroch~[\GeroPhD]):

\Theorem
Let ${\cal M}$ be a manifold that interpolates between a surface
${\cal S}_1$ and a surface~${\cal S}_2$. Suppose that ${\cal M}$
admits a smooth vector field $V^a$, every integral curve of which has an
endpoint on~${\cal S}_1$. Then ${\cal S}_1$ is connected.
(And similarly for ${\cal S}_2$.)

\Proof
Suppose that
${\cal S}_1$ is not connected and suppose, further, that it has two disjoint
components ${\cal A}_1$ and ${\cal A}_2$ (which may not themselves be
connected). The manifold
${\cal M}$ can then be decomposed into two disjoint
sets ${\cal N}_1$ and ${\cal N}_2$, where
$$
{\cal N}_i=\{p\mid \hbox{the integral
curve of $V^a$ through $p$ intersects\ } {\cal A}_i\}.
$$
Both these sets
will be open. This is not possible, since $\cal M$ is connected.
\EndProof

\Section The Dynamics of Topology Change

Theorem~1 does not completely rule out classical topology change.
For, there are solutions of Einstein's equation (with reasonable sources)
that have causality violations~-- the G\"odel, Kerr and Taub-NUT solutions,
among others (ref.~[\HE], p.~161--178).
It is true that such scenarios are difficult
to interpret, or to reconcile with our experience. But, allowing topology
change to occur opens up the possibility of such rich spacetime structures,
that it is tempting to try anyway. (It might be possible, for example,
to construct an interpretational framework in which only the initial
and final states are viewed as physically ``real,'' with the interpolating
spacetime regarded as a device for carrying out calculations~-- as in the
usual interpretation of the
path integral approach to quantum theory.) The obvious question is, does
Einstein's equation constrain topology change in any way, when the source
is restricted in some reasonable manner?
(If the source is not restricted, Einstein's equation provides no
constraint at all, for then any metric is a solution.)
The result that I discuss below suggests that Einstein's equation does place
significant constraints:
with certain restrictions on the curvature (which mainly follow
from reasonable restrictions on the source), it appears that topology
change is not allowed for causally compact spacetimes. This result,
but with slightly stronger assumptions, was obtained
previously by Tipler~[\TiplPhD\,\Tipl] for the closed universe and
externally Lorentzian cases.

\Theorem
Let $\cal M$ be a (time-oriented) causally compact
interpolating spacetime of dimension $\geq 3$
with initial surface ${\cal S}_1$ and final surface
${\cal S}_2$. Suppose that
\item{i)} every full (i.e., no endpoint on ${\cal S}_1$ or on ${\cal S}_2$)
null geodesic has a point on it at which
$F_{abcd} \equiv U_{[a}R_{b]ef[c}U_{d]}U^eU^f \not= 0$, where $U^a$~is the
tangent to the geodesic; and
\item{ii)} for any point~$u_0$ on a past-complete,
affinely parametrized null geodesic~$\lambda (u)$,
the half-integral null convergence condition (explained below) holds
along~$\lambda$ to the past of~$u_0$.
\ContinueThm
Then $\cal M$ is diffeomorphic to ${\cal S}_1\times[0,\,1]$, and ${\cal S}_1$
and ${\cal S}_2$ are connected.

It is important to note that this result is not a ``singularity theorem,''
at least of the standard type, i.e., {\it it makes no assumptions about
geodesic completeness}. As it turns out, the geodesics that
we are interested in here are guaranteed to be complete (in the direction
of interest) under the
conditions of the theorem (this is a standard result~-- see lemma~A
below). Thus the theorem does not allow topology
change to be had, even at the price of a (standard) singularity.
The question of possible singularities associated with topology
change~-- when one of the conditions imposed above is relaxed~-- is
discussed again in section~1X.

Two restrictions are made on the curvature in the theorem. I discuss
each in turn:

\SubSection Assumption i

This assumption is essentially a generality condition; i.e., it will fail
to hold only in situations that are, in a precise sense, highly special.
There are various ways in which such
generality assumptions may be made. The condition that I am using is the
null form of the ``generic condition'' that was
used by Hawking and Penrose in their 1970 singularity theorem
\Refc{S.W. Hawking and R. Penrose, \Jou{Proc. Roy. Soc. Lond.}, \Vol{A314},
529 (1970).}\HE].
Their original argument in support of the reasonableness of
the condition was made in four dimensions, but similar arguments may
be made in higher dimensions as well
\StoreRef{\BH}
\Ref{J.K. Beem and S.G. Harris, \Jou{Gen. Rel. and Grav.}, \Vol{25},
939 (1993); \Jou{Gen. Rel. and Grav.}, \Vol{25}, 963 (1993)}.

First, consider non-vacuum spacetimes (i.e., $T_{ab} \ne 0$).
For null vectors $U^a$, Einstein's equation implies that
$R_{ab}U^aU^b = kT_{ab}U^aU^b$, where~$k$ is a constant (the relationship
holds even if there is a cosmological constant).
For matter energy-momentum tensors of known types,
$T_{ab}U^aU^b$ will vanish only if $T_{ab}$ represents pure radiation
travelling in the $U^a$ direction (ref.~[\HE], p.~101).
In any realistic model with sources
it is reasonable to suppose that there are other types of matter in addition
to pure radiation, or that a null geodesic does not align itself with the flow
of radiation throughout, i.e., it is reasonable to require that
$R_{ab}U^aU^b\not= 0$ somewhere on each null geodesic. At that point
we must also have $F_{abcd}\not= 0$.

In the vacuum case (i.e., ``pure gravity'') in spacetimes of dimension
greater than three,
the null generic condition is equivalent
to requiring that $U^a$ not point in a principal null direction of the Weyl
tensor throughout the length of the null geodesic. Since at any point there
are only a finite number of such directions~[\BH],
it is reasonable to require that such an alignment not occur.

In three dimensions
there is no vacuum case, for $C_{abcd}\equiv 0$ and the
Riemann tensor is determined by the Ricci tensor.
But there is further reason here to believe that $R_{ab}U^aU^b$
will not generically vanish along the entire length of a geodesic.
Consider the expression
$R_{ab}U^aU^b$, for all null vectors $U^a$ at a point. Pick a basis
$\{ P^a,\; M^a,\; S^a\}$ at that point such that $P^a$ and $M^a$ are
both null, $S^a$ is a unit spacelike vector orthogonal to $P^a$ and $M^a$,
and $P^aM_a=-1$. Then an arbitrary null vector is either
proportional to $M^a$ or is proportional to a vector of
the type $P^a+(\beta^2/2)M^a+\beta S^a$. Thus, $R_{ab}U^aU^b=0$ will
yield a quartic equation for $\beta$, unless $R_{ab}M^aM^b$ is already zero
whereupon the equation is cubic; i.e., there are at most 4 null directions,
$U^a$, at a point that can satisfy $R_{ab}U^aU^b=0$. And, as above, it seems
reasonable to suppose that in a generic spacetime a null geodesic will
not align itself throughout along one of these directions; i.e.,
that $R_{ab}U^aU^b$, and therefore $F_{abcd}$, is non-zero somewhere.

A stronger form of the generic condition that we might consider using requires
that $F_{abcd} \not= 0$ somewhere on one side of a given point on every
geodesic. This form is useful in situations where we are interested
in what happens to the future (or has happened to the past) of some initial
(or final) state. In gravitational collapse, for example, we are interested
in what happens
to the future of the initial state. And, in discussions of the initial
singularity, we are interested in the past. In such situations it is often
necessary to impose conditions on the half-geodesics that have initial
(final) endpoints at the initial (final) state. If this is not done,
we often lose important information. For example,
the Hawking-Penrose singularity
theorem does not make assumptions about half-geodesics and, as a
consequence, it does
not yield information about the location (in time) of the singularity.
In fact, it leaves open the possibility that the singularity of
gravitational collapse might have occurred to the past of the (non-singular)
initial state, or that the `initial' cosmological
singularity might occur in the future.
On the other hand, with a condition on half-geodesics
it is possible to ensure, for instance, that
the initial singularity does indeed lie to the past
\Ref{A. Borde, \Jou{Class. and Quant. Grav.}, \Vol{2}, 589 (1985).}.

The problem with assumptions on half geodesics, however, is that they are
considerably stronger than
than conditions on full geodesics. Consider a situation where a black hole
forms out of some realistic (and therefore asymmetrical) initial state.
It will settle down to a highly symmetrical final state~[\HE]. In such a case
it is much weaker to require that $F_{abcd} \not= 0$ somewhere on a geodesic
than to require, say, that this happens at some point after the black hole
forms.

Topology change is a problem where we are chiefly interested in that
portion of a full spacetime that lies between the initial and final
surfaces, the interpolating spacetime~$\cal M$. It might seem that to extract
information we must require that every geodesic in~$\cal M$ have a point at
which $F_{abcd}\not= 0$, even if the geodesic has endpoints
on~$\partial{\cal M}$.
Such a condition was imposed in Tipler's theorem~[\TiplPhD\,\Tipl].
But,
as we shall see, the weaker condition that $F_{abcd}\not= 0$ somewhere just
on full geodesics is enough.

\SubSection Assumption ii

Let $\lambda (u)$ be an affinely parametrized null geodesic, where
$u$~increases in the future direction.
The {\it half-integral null convergence condition\/} is said to hold
along~$\lambda$ to the past of some point~$u_0$ (more precisely, to
the past of~$\lambda(u_0)$) if for any $\delta > 0$,
$\exists b > 0$ such that for any $u_1 < u_0$ there is an interval
$I$ of length $\geq b$, with $u_1 >\sup I$ and with
$$
\int_{u}^{u_0}R_{ab}U^aU^b\,du \geq -\delta,
\qquad \forall u\in I.
$$

This condition limits how negative the ``matter term'' $R_{ab}U^aU^b$
can get\StoreRef{\Borde}
\Ref{A. Borde, \Jou{Class. and Quant. Grav.}, \Vol{4}, 343 (1987).}.
(The quantity $R_{ab}U^aU^b$ is
called the matter term because it may be related, via Einstein's
equation, to $T_{ab}U^aU^b$.)
The precise statement of the condition is, unfortunately,
a little involved, but the idea that it tries to express is simple. Suppose
that $\lambda (u)$ is a null geodesic with affine parameter~$u$ and tangent
$U^a$. Let $p=\lambda (u_0)$ be some point on the geodesic.
Now, suppose that in some places along~$\lambda$ before~$p$
the term $R_{ab}U^aU^b$ is negative. The condition then requires that there
always be other regions where this term is positive enough to make
$\int_{u}^{u_0}R_{ab}U^aU^b du$ come out as close to zero as we
want, for all~$u$ in an interval of some finite length~$b$. (Roughly
speaking, it asks that $\int \!R_{ab}U^aU^b du$ at least come repeatedly
close to zero as one looks along~$\lambda$ in the direction of
decreasing~$u$.) This condition is a more stringent
restriction on the curvature than the one in assumption~(i). It will be
discussed in greater detail in the next section.

\SubSection The proof

The proof of Theorem~3 depends on observing that there must be a non-empty
future Cauchy horizon $H^+({\cal S}_1)$, if ${\cal S}_1$ and ${\cal S}_2$
are not diffeomorphic.
It is a standard result in global general relativity that
such horizons are ``generated by null geodesics'' (in a sense that is
defined below). The main thrust of the proof
involves first showing that $H^+({\cal S}_1)$ contains a full geodesic, and
then observing that focusing effects that occur on this null geodesic,
as a consequence of assumptions~(i)
and~(ii), are incompatible with certain consequences of the assumption
of causal compactness.

The detailed proof relies on some
standard results from global general relativity.
Let ${\cal H}=H^+(\cal S)$ be a future Cauchy
horizon, where $\cal S$ is a spacelike surface,
and let ${\cal A}=\dot I^-(p)$ be the boundary of the past of $p$, where
$p$ is any point.
Then the following hold (ref.~[\HE], chapter~6):

\Bull{(a)\/}
$\cal H$ and $\cal A$ are achronal, i.e., no two points on either of
these sets can be connected by a timelike curve.

\Bull{(b)\/}
Through each point of $\cal H$ there passes a
past-directed null geodesic that either does not leave $\cal H$ when
followed into the past or leaves it at the edge of the surface~$\cal S$.
This geodesic may leave $\cal H$ at any point when followed into the future.
(See fig.~2c and fig.~2d for an illustration of this behavior.)

\Bull{(c)\/}
Through each point of $\cal A$ there passes a future-directed null
geodesic that can leave $\cal A$, when followed into the future, only
through $p$, but may leave $\cal A$ anywhere when followed into the past.

\Bull{(d)\/}
let $\mu$ and $\rho$ be
future-directed null geodesics that intersect at some point~$q$;~if $q'$
lies to the future of~$q$ on~$\mu$, or if $q'\in I^+(q)$, then
the portion of~$\rho$ to the past of~$q$ lies in~$I^-(q')$.
Similarly, if the future endpoint of a timelike or null curve~$\rho$
coincides with the past endpoint of another timelike or null
curve~$\mu$, with a discontinuity in their tangents at the intersection
point (i.e., they ``meet at an angle''), then $\rho \subset I^-(\mu)$.

\smallskip
The null geodesics in (b) and (c) are called the
{\it null generators\/} of $\cal H$ and~$\cal A$.
Before embarking on the main proof, it is useful to state two
standard results that we will need:

\Declare{Lemma A} Let~$\cal H$ be a future Cauchy horizon and~$\lambda$
a generator of this horizon with no past endpoint on~$\cal H$.
If~$\lambda$ is confined within a compact set when
followed into the past, then~$\lambda$ is past
complete\/~{\rm (ref.~[\HE], p.~295--297, and ref.~[\TiplPhD], p.~98--103)}.

It is this
result that makes it unnecessary for us to assume geodesic completeness
in~\PrevThm.

The second result describes the focusing of null geodesics in an
$n$-dimensional spacetime ($n>2$).
The standard four-dimensional procedure is simply
imitated: Consider a congruence of null geodesics; let $u$~be an affine
parameter along the null geodesics, chosen to increase in the past direction,
and let $U^a=({\partial/\partial u})^a$. Let $D_a$ be the covariant
derivative, and let $D=U^aD_a$. At each point set up a basis of vectors
consisting of $U^a$, a null vector $M^a$ obeying $M^aU_a=-1$ and $DM^a=0$,
and $(n-2)$ spacelike vectors $S^a_i$ orthogonal to each other and to
$U^a$ and $M^a$ and also obeying $DS^a_i=0$. Then $h_{ab}=g_{ab}+
2U_{(a}M_{b)}$ will be a positive-definite metric on the space orthogonal
to $U^a$ and $M^a$. Define the expansion of the congruence by $\theta=
D_aU^a$ ${(= h^{ab}D_aU_b}$) and the shear by $\sigma_{ab}=D_{(a}U_{b)} -
{1\over n-2}h_{ab}\theta$. Let $\sigma_{ij}=S^a_iS^b_j\sigma_{ab}$. Then,
$2\sigma^2\equiv\sigma_{ab}\sigma^{ab}=\sigma_{ij}\sigma^{ij}\geq0$, with
equality holding iff $\sigma_{ij}=0$.
When the spacetime dimension~$n$ is greater than~$3$,
the behavior of $\theta$ and
$\sigma_{ij}$ along the null geodesics of the congruence is given by:
$$
\displaylines{
\hfill{d\theta\over du} = -{1\over n-2}\theta^2 -
R_{ab}U^aU^b - 2\sigma^2\hfill(1)\cr
\hfill{d\sigma_{ij}\over du} = -C_{iajb}U^aU^b-
{2\over n-2}\theta\sigma_{ij}-\sigma_{ik}\sigma_{jl}h^{kl}+{2\sigma^2\over
n-2}h_{ij}\hfill(2)\cr}
$$
where $C_{abcd}$ is the Weyl tensor. When~$n=3$, $\sigma_{ij}\equiv 0$
and we have
$$
{d\theta\over du} = -\theta^2 - R_{ab}U^aU^b.\eqno{(1')}
$$
In both cases
we are interested in focusing,
i.e., in the conditions under which~$\theta$ diverges.

\Declare{Lemma B} Suppose that the expansion~$\theta$ is positive at
some point~$\lambda(u_0)$ of a member~$\lambda$ of a congruence of
past-directed null geodesics (with affine parameter~$u$ chosen to
decrease in the past direction).
Further, suppose that $\lambda$ is past-complete and that the
half-integral null convergence condition holds to the
past of~$u_0$. Then, $\theta\rightarrow \infty$
within a finite affine parameter distance to the past of~$u_0$~{\rm[\Borde]}.

Here, now, is the proof of~\PrevThm:

\Proof
The main argument is best given in steps:

\Heading{1. There is a non-empty future Cauchy horizon~$\cal H$}
Set up a timelike vector field $V^a$ on $\cal M$, and suppose that one of its
integral curves does not intersect either ${\cal S}_1$ or ${\cal S}_2$. Then,
as we have seen in Theorem~1, there will be a closed timelike curve in
$\cal M$.
This curve cannot intersect ${\cal S}_1$, and so points on it cannot lie in
$D^+({\cal S}_1)$. Therefore ${\cal H}=H^+({\cal S}_1)\not=\emptyset$.
(It is assumed here that~${\cal S}_1$ is connected. If it is not, an
argument identical to the one given below may be applied to a connected
component of~${\cal S}_1$, with an identical ensuing contradiction.)

Since I am allowing interpolating spacetimes that are confined
within a timelike tube~$\cal T$,
there is a slight subtlety here. The
interpolating spacetime~$\cal M$~-- bounded by ${\cal S}_1$, ${\cal S}_2$
and ${\cal T}$~-- is the region of interest.
Thus a point~$p$ will lie in
$D^+({\cal S}_1)$ as long as every past-directed timelike or null curve
from it
{\it that is confined to~$\cal M$} eventually intersects~${\cal S}_1$.
Now, let~$\gamma$ be a past-directed timelike or null curve
and suppose that $\gamma$ intersects~$\cal T$. As long as
$\gamma$ can be continued~-- either along~$\cal T$ or within the interior
of~$\cal M$~-- to~${\cal S}_1$, it is considered a curve that
meets~${\cal S}_1$.

\Heading{2. The generators of $\cal H$ are past complete}
It follows from the comment in the preceding paragraph that no
null generator of~$\cal H$ can intersect~$\cal T$ when followed into
the past. For, suppose there is a generator~$\nu$ of~$\cal H$
that intersects~$\cal T$ at a past endpoint~$x$. Let~$\mu$ be
any timelike curve on~$\cal T$ with~$x$ as its past endpoint,
and let~$x_i$ be a sequence of points
on~$\mu$ converging to~$x$. Each point~$x_i$ will
lie to the future of some point on~$\nu$, and thus none of the~$x_i$
will lie in~$D^+({\cal S}_1)$. So from each~$x_i$ there
will be a past-directed timelike curve~$\alpha_i$ that never
intersects~${\cal S}_1$ when followed into the past within~$\cal M$.
The curves~$\alpha_i$ are thus trapped in~$\cal M$.
The sequence
$\{\alpha_i\}$ of these curves will have a past-directed limit
curve, $\alpha$, through~$x$ (ref.~[\HE], p.~185).
The curve~$\alpha$ must also lie in~$\cal M$.
Points on~$\alpha$ close to~$x$ must lie to the
chronological past of points on~$\nu$ (since $\alpha$ will meet~$\nu$
``at an angle'' at~$x$) and thus they will lie
in the interior of~$D^+({\cal S}_1)$.
But this will make points on some of the
curves~$\alpha_i$ also lie in~$D^+({\cal S}_1)$, contradicting the
fact that the~$\alpha_i$ have been chosen not to
intersect~${\cal S}_1$.

Now, let $p$ be any point in the interior of $\cal M$ that lies on a null
generator of $\cal H$ and is not its future endpoint, and
let $\lambda$ be the portion of this null geodesic that lies to
the past of $p$. Let $p'$ be some point in $I^+(p)$. Then $\lambda \subset
\overline{I^-(p')}$, i.e., it is a subset of a compact set
(by causal compactness of $\cal M$). It follows from lemma~A that
$\lambda$ is necessarily past complete.
Clearly, this applies to any null generator of~$\cal H$.

\Heading{3. $\cal H$ contains a full geodesic}
Let~$p$, $p'$ and~$\lambda$ be, respectively, the points and the curve
from step~2, and let ${\cal B}={\cal H}\cap\overline{I^-(p')}$.
When~$\lambda$ is extended into the future it may or may not leave
${\cal B}$. Suppose it does leave~$\cal B$ at some future endpoint.
Then there will be some other null generator of $\cal H$ that does
not leave $\cal B$ to the future. To see this, let $\{ q_i\}$ be a
sequence of
points on $\lambda$ such that each $q_{i+1}$ lies to the past of $q_{i}$,
and with no finite segment of $\lambda$ containing an infinite number of
these points. These points will have a limit point~$q$. Now,
$\lambda\subset \cal B$. Also $\lambda\subset \dot I^-(p)$. Both these
sets are closed and so the point $q$ belongs to both. The null generator
of $\cal B$ through $q$ when followed into the future will also lie on
$\dot I^-(p)$. Call this generator $\mu$.
I first show that~$\mu$ cannot represent the same geodesic as~$\lambda$
itself.

Suppose, to the contrary, that it does. Choose a point~$x$ on the
future extension of~$\lambda$, after it leaves the set~$\cal B$.
Let~$\cal O$ be
a neighborhood of the segment of~$\lambda$ between~$x$ and~$q$ such that
there is some point~$\hat q$ on~$\lambda$ to the past of~$q$ that does
not lie in~$\cal O$.
{}From now on let the sequence $\{ q_i\}$ be restricted to those
points that lie to the past of~$\hat q$. Let
$N_{y}$ represent the tangent to~$\lambda$ at any point~$y\in\lambda$.
Since~$q_i\to q$ we must have $N_{q_i}\to N_q$ (otherwise there
will be some other generator of~$\cal B$ that intersects~$\lambda$ at~$q$,
and that is not possible).
This means
that from all the~$q_i$ sufficiently close to~$q$ the null geodesics with
initial tangents~$N_{q_i}$ must reach an arbitrarily small neighborhood
of~$x$ without ever leaving~$\cal O$. This is not possible since
(i)~all of~$\lambda$ to the past of~$q$ lies in~$\cal B$,
(ii)~$x\notin{\cal B}$, and
(iii)~the segment of~$\lambda$ between each~$q_i$ and~$q$ contains the
point~$\hat q$ that does not lie in~$\cal O$.

So~$\mu$ cannot represent the same geodesic as~$\lambda$.
It (i.e., $\mu$) can leave $\dot I^-(p)$ to the
future only through $p$. Suppose it does. But then for any point~$b$
in $\cal H$ to
the future of $p$ along~$\lambda$, we have $\mu\subset I^-(b)$, which is
not
possible since $q\in\mu$, and $q$ and~$b$, being points on~$\cal H$, cannot
be connected by a timelike curve. So, $\mu$ does not leave
$\dot I^-(p)$ to the future, and therefore it does not leave
$\overline{I^-(p')}$ to
the future either. Also, it does not leave $\cal H$ to the future.
For, suppose it does. Let $c$~be a point on~$\mu$
such that  $c\in\dot I^-(p)$ but $c\notin \cal H$.
Points close to~$c$ will not lie in
$D^+({\cal S}_1)$ but some will lie in $I^-(p)$. Let $\rho$ be a
future-directed
timelike curve from one of these points to~$p$. Points on $\rho$ close to
$p$ will lie in $D^+({\cal S}_1)$, i.e., $\rho$ must enter
$D^+({\cal S}_1)$ at some point on $\cal H$ before it reaches~$p$. This
violates the achronal nature of $\cal H$. So, $\mu$ does not leave $\cal H$
to the future. Thus, it does not leave $\cal B$ to the future or to the
past (since it cannot leave either $\cal H$ or $I^-(p')$ to the past).
(A similar result was obtained previously in a different way
\Ref{A. Borde, \Jou{Phys. Lett.}, \Vol{102A}, 224 (1984).}.)

\Heading{4. Focusing occurs on $\cal H$, leading to a contradiction}
Next, we consider the focusing of null geodesics in $\cal B$. To do
this, we need to define a congruence of null geodesics in a neighborhood
of $\cal B$ (so that quantities like derivatives may be defined). This can
be done by varying points on $\cal B$ in a direction not contained
in~$\cal B$~[\Borde].
All the quantities that we are interested in ($\theta$, etc.) turn out to
be independent of the particular variation that is used. Choose an
affine parameter on the geodesics of this congruence that increases
in the future direction.

Since $\cal B$ is compact, it may be shown that we cannot have
$\theta\leq 0$
throughout and ${\theta< 0}$ somewhere (ref.~[\HE], p.~297--298).
Therefore, we either have
$\theta = 0$ throughout, or $\theta > 0$ somewhere on $\cal B$.
If the spacetime dimension~$n$ is~3, assumption~(i) of the theorem
means that
$R_{ab}U^aU^b$~-- and hence from equation~(1$'$), $\theta$~-- cannot be
zero everywhere along~$\mu$.
If $n>3$ and if
$\theta$ is zero throughout on $\mu$, then equation~(1) will yield
$R_{ab}U^aU^b=-2\sigma^2\leq 0$. From assumption~(ii) of the theorem,
it follows that $R_{ab}U^aU^b=0$ throughout; therefore $\sigma_{ij} = 0$
throughout as well. But, from assumption~(i) and equation~(2)
applied to $\mu$, it follows that $\sigma_{ij} \not= 0$ somewhere on $\mu$.
So, we must have $\theta > 0$ somewhere on $\mu$. From lemma~B,
$\theta\rightarrow \infty$
within a finite affine parameter distance.
This will violate the achronality of the horizon (ref.~[\HE], p.~115--116).
(Just as the crossing of two
separate generators of~$\cal H$ will violate the achronality of~$\cal H$,
by the
time reverse of (d)~above, so also will the crossing of `infinitesimally
close' generators, the diverging of~$\theta$ being taken to indicate
such a crossing.)

Therefore, there can be no Cauchy horizon under the
assumptions of the theorem, i.e., every integral curve of~$V^a$ must take on
endpoints on ${\cal S}_1$ and on ${\cal S}_2$. Hence the first part of
the result.

\smallskip
The second part follows immediately from these facts: (a)~${\cal M}$ is
connected, and (b)~${\cal M}=[0,\,1]\times {\cal S}_1$ (as shown above).
\EndProof

\Section Energy conditions

Assume that Einstein's equation holds (possibly with cosmological
constant) on $\cal M$. What sorts of conditions on the matter energy-momentum
tensor, $T_{ab}$, will yield the half-integral null convergence condition
of the previous theorem?

Obviously, any condition on $T_{ab}$ that implies that $R_{ab}U^aU^b\geq 0$
(this is called the {\it null convergence condition\/}) will be sufficient.
An important case is the vacuum: $T_{ab}=0$. This covers those situations
where we are interested in the behaviour of ``pure gravity.'' Such
situations include models where ``matter'' is built out of gravity either
by using non-trivial topologies~[\WhTwo\,\MisWh\,\FrSork], or higher
dimensions, or both~[\SorkTwo\,
\Refe{R. Sorkin, \Jou{Phys. Rev. Lett.}, \Vol{51}, 87 (1983); {\bf 54},
86(E) (1985).}.

Another case of interest is when $T_{ab}\not= 0$ but is bounded below
in some suitable sense. For example, suppose at some $p\in\cal M$ that
$T_{ab}V^aV^b \geq K$ for all unit timelike $V^a$ at~$p$, where~$K$
is some number (possibly negative). Physically this means that the
energy density as seen by any observer passing through~$p$ will
be $\geq K$. It follows that $R_{ab}U^aU^b\geq 0$ for all null
$U^a$ at $p$. For, let $U^a$ be any null vector at~$p$ and
let $V^a_i$ be a sequence of unit timelike vectors at~$p$ that approaches
the direction of~$U^a$
as a limit direction. Let $T^a_i = b_iV^a_i$ (no sum on $i$) such that
the vectors $T^a_i$ converge to $U^a$. Then we must have $b_i\rightarrow 0$.
Therefore, $T_{ab}T^a_iT^b_i = T_{ab}V^a_iV^b_i(b_i)^2\geq K(b_i)^2
\rightarrow 0$. Thus $T_{ab}U^aU^b\geq 0$ and hence, by Einstein's equation,
$R_{ab}U^aU^b\geq 0$. (This result was proved previously by Tipler for
a special class of energy-momentum tensors~[\TiplPhD\,
\StoreRef{\TiplAWEC}
\Refe{F.J. Tipler, \Jou{Phys. Rev.}, \Vol{D17}, 2521 (1978);
\Jou{J. of Diff. Eqns}, \Vol{30}, 165 (1978).}.)

Though in this result $K$~can be negative, and can vary from point to
point on~$\cal M$, the physically interesting case appears to be $K=0$. In
this case we say that $T_{ab}$ obeys the {\it weak energy condition}, i.e.,
all observers measure a non-negative energy density. This condition is
obeyed by all known forms of classical matter (ref.~[\HE], p.~89--91).

The energy-momentum tensor associated with quantum fields can, however,
have expectation values that violate the weak energy
condition~[\TiplAWEC\,
\StoreRef{\TomRom}
\Refm{T.A. Roman, \Jou{Phys. Rev.}, \Vol{D33}, 3526 (1986);
\Jou{Phys. Rev.}, \Vol{D37}, 546 (1988).}\MTY].
Can such fields be used to drive topology change so that the process, though
forbidden classically, can occur semi-classically? When dealing with such
situations, the freedom that we have in allowing $K$ to be negative is not
likely to be of much use. Suppose that there is a lump of
good (i.e., positive energy) matter somewhere: then all observers will measure
the energy to be positive. The value of this measured energy will
not be bounded above, for by moving past the lump at speeds approaching the
speed of light the energy can be made to appear as large as we wish.
Similarly for bad matter then, if the energy as measured by some observer at
a point $p$ is negative, it seems unreasonable to expect in general
that there be a lower bound for the value of the measured energy for all
observers passing through $p$. (Even though, formally, it is possible to
construct examples where this happens: a simple one is $T_{ab}=Cg_{ab}$.) This
is born out in calculations of the energy-momentum tensor that have
been done for quantum fields. For example, for the Casimir effect
or for Hawking radiation the violations of the weak energy condition
appear not to be bounded below~[\TomRom].

In such cases, $R_{ab}U^aU^b$ can be negative. If the regions where this
happens do not dominate the spacetime then we might expect that on the
average geodesics see non-negative $R_{ab}U^aU^b$, i.e., that integral
conditions such as the half-integral convergence condition might perhaps
hold.
This is so far only a piece of wishful thinking. Integral convergence
conditions were introduced by Tipler~[\TiplAWEC] to replace the standard
pointwise convergence conditions in studies of geodesic focusing. Though
much further work has been done on such conditions~[\Borde\,\TomRom\,
\Refm{C. Chicone and P. Ehrlich, \Jou{Manuscripta Math.}, \Vol{31},
297 (1980).}
\Refm{U. Yurtsever, \Jou{Class. Quant. Grav.}, \Vol{7}, L251 (1990).}
\Refm{G. Klinkhammer, \Jou{Phys. Rev.}, \Vol{D43}, 2542 (1991).}
\Refm{R. Wald and U. Yurtsever, \Jou{Phys. Rev.}, \Vol{D44}, 403 (1991).}
\Refe{L. Ford and T. Roman, Tufts University preprint (1994).}
their ultimate usefulness remains uncertain.
The question needs further study.

An indication that we might be able to make some reasonably reliable
statement about the extent of the violations of the weak energy condition
when quantum fields are present comes from a study of black hole
evaporation
\Ref{S.W. Hawking, \Jou{Comm. in Math. Phys.}, \Vol{43}, 199 (1975).}.
The qualitative picture of some aspects of this process
that we are led to believe (from using energy conservation to relate the
energy flux at large radial distances to the energy flux near the black
hole)
\Ref{J. Bardeen, \Jou{Phys. Rev. Lett.}, \Vol{46}, 382 (1981).}
is this: Consider a
Schwarzschild black hole of initial mass~$m$. If it were not evaporating,
the null geodesic generators of its event horizon would sit at $r=2m$
throughout (assuming that no new matter falls in).
During evaporation, however, these null geodesics that
were initially at $r=2m$ will start diverging and will escape to infinity
(because, crudely, the black hole mass decreases and the forces holding
them in place get weaker). Slightly inside $r=2m$ (at $r\sim 2m(1-(m_p/m)^2)$,
where $m_p$ is the Planck mass), there will be a null
hypersurface whose generators were initially converging slightly. This
convergence will slowly go to zero and the surface will be, by definition,
the event horizon. Inside it there will be null geodesics that were also
initially converging and which continue to do so with the convergence
not approaching zero. On all these geodesics the violations of the weak
energy condition cannot be uncontrollably large~-- for, in that case, one
would expect the initially converging geodesics slightly inside
$r=2m$ to also
all quickly diverge to infinity. All the preliminary calculations suggest that
this does not happen~-- offering the hope that some sensible estimate
can be made, both of the extent of the violations of the weak energy
condition in this quantum process (and in others), as well as of the effect
that such violations might have. It might be possible, for example, that
a singularity theorem might be provable for evaporating black holes
\Refc{R. Penrose, unpublished remark.}\TomRom\,
\Refe{L. Ford and T. Roman, Tufts University preprint (1994).}
using the idea that violations of the weak energy condition are limited.
And, it might be possible to make some statement about topology change,
even in the presence of quantum fields. The statement of assumption~(ii)
as an integral convergence condition is an attempt to provide a framework
for doing this~-- but, clearly, a lot more remains to be done.

\Section Time-orientability

Would the conclusions of this paper be significantly different
were we to drop our insistence on time-orientable metrics?
If the metric on $\cal M$
is not required to be time-orientable, then
some changes do show up at the kinematical level.
For example, Sorkin has shown that the Reinhart-Sorkin selection rule,
$\Delta\chi \equiv \chi({\cal S}_1) -\chi({\cal S}_2) = 0$, is no longer a
condition for the existence of a Lorentzian cobordism in odd dimensions
\Ref{R. Sorkin, \Jou{Int. J. of Theor. Phys.}, \Vol{25}, 877 (1986).}.
The results of this paper are, however, still essentially true.
To see this, note
first that the existence of a not-necessarily-time-orientable metric is
equivalent to the existence of a (timelike) direction field
(i.e., at each point
a vector is defined up to sign)~[\HE]. Integral curves to this field may
still be constructed (if necessary, first in the time-orientable
double covering manifold~(ref.~[\HE], p.~181) and then mapped back to the
original manifold). If the interpolating spacetime is assumed to be
causally compact, it follows that there will be a closed timelike
curve unless each of these integral curves takes on two endpoints
on $\partial\cal M$.
(The existence of the closed timelike curve may first be proved,
if necessary, in the time-orientable double covering manifold; this curve
may then be mapped into a closed timelike curve in~$\cal M$.)
Therefore, if causality violations are to be excluded,
$\cal M$ may be divided into three disjoint sets, ${\cal N}_1$, ${\cal N}_2$
and $\cal N$, where
$$
{\cal N}=\{p\mid \hbox{\sl the integral curve through $p$ takes
on one endpoint each on ${\cal S}_1$ and ${\cal S}_2$}\},
$$
and
$$
{\cal N}_i=\{p\mid \hbox{\sl the integral curve through $p$ takes on both
endpoints on\ } {\cal S}_i\}.
$$
These are all open sets, and so, since
$\cal M$ is connected, two of them must be empty. Therefore, $\cal M$ has
either a single boundary ${\cal S}_1$
(or ${\cal S}_2$), which is mapped onto itself by the integral curves
of the direction field, or it has two
separate, but diffeomorphic, boundaries. In neither case can topology
change be said to occur. (A proof along these lines was given
by Geroch for the case of closed and externally Lorentzian
spacetimes~[\GeroPhD]. Geroch's result is sometimes quoted as having
assumed that ${\cal S}_1$ and ${\cal S}_2$ are connected. In fact, the
proof works even if they are not, as long as $\cal M$ is connected.)
An example of the first case is the M\"obius strip with time along the
non-circular direction and the space axis along the circular direction.
Here, the single $S^1$ boundary is mapped to itself.
See \NFig.
In the second case, when ${\cal S}_1$
and ${\cal S}_2$ are diffeomorphic (and non-empty), a time-orientation
may in fact be introduced on $\cal M$ as follows: choose a parameter~$t$
along the integral curves that increases from ${\cal S}_1$ to
${\cal S}_2$, then pick $({\partial/\partial t})^a$ to be the
future direction.
\StartFigure{80}{60}
          {newpath 0 0 moveto
           -100 0 translate
           gsave
           newpath
           1 .5 scale
           1.3 setlinewidth
           140 100 40 -90 180 arc stroke
           newpath
           140 40 40 -120 0 arc stroke
           newpath
           140 40 40 55 153 arc stroke
           newpath
           140 40 40 170 200 arc stroke
           grestore
           newpath
           100 50 moveto
           100 35 100 8 120 3 curveto stroke
           newpath
           108 15 moveto
           120 30 130 30 140 30 curveto stroke
           /Vfield {newpath
           140 3 moveto
           143 6 lineto 140 3 lineto 137 6 lineto
           140 3 lineto 140 27 lineto
           143 24 lineto 140 27 lineto 137 24 lineto
           gsave .7 setlinewidth stroke grestore} def
           Vfield
           gsave 0 40 translate Vfield grestore
           newpath
           180 20 moveto 180 50 lineto [3 3] 0 setdash stroke
         }
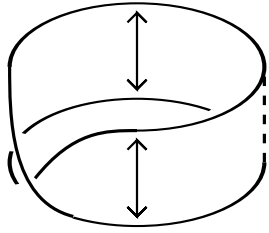
\caption{A non-time-orientable ``transformation'' of a circle onto itself
in a M\"obius strip spacetime. Compare this case
with the M\"obius strip example of fig.~6c; that example may be viewed
as a $\emptyset\leftrightarrow S^1$ transition whereas the
non-time-orientability
of this one makes it difficult to interpret.}
\EndFigure

Thus, we have the following result: a causally compact
spacetime that interpolates between non-empty boundaries ${\cal S}_1$ and
${\cal S}_2$ and which
does not have closed timelike curves must (a) be diffeomorphic to
${\cal S}_1 \times [0,\,1]$, and (b) hence be time-orientable.

Therefore, topology change will have causality violations associated
with it, even in the non-time-orientable case.
Further, theorem~3 may be applied in the
time-orientable double covering manifold
(all the other assumptions of the theorem will continue to hold)
to get a contradiction.

No other orientability requirements were placed on the interpolating
spacetime. If the manifold itself is orientable, but the metric on
it is not time-orientable, then it cannot be space-orientable either~[\GH].
It might be that violations of~$P$ and of~$CP$ in particle interactions
force us to only consider orientable manifolds as models for space. But,
if this is true, and if we wish to consider non-time-orientable
metrics (despite the implication of the $CPT$ theorem that
$T$ must be violated if $CP$ is), we can do so by looking
at non-orientable interpolating manifolds. Thus there appears to be
no overwhelming reason against formally considering non-time-orientable
metrics~-- but, as we have seen above, nothing of overwhelming
significance seems to emerge either from doing so.

\Section A Few Words on Differentiability

I have assumed in the discussion so far that the manifolds under discussion
are smooth and that the various fields defined on them are smooth as well.
In fact, we usually need a much lower degree of differentiability.
In order to discuss this, the following standard notation (already used
in this paper) is helpful: $C^n$ means
$n$ times differentiable, with continuous $n$-th derivatives; {\it smooth\/}
means $C^{\infty}$, i.e., all derivatives exist.
Now, integral curves of the vector field $V^a$ will exist
if $V^a$ is continuous, and the curves will be unique if $V^a$ is~$C^1$
(this is a
sufficient condition; the necessary condition is weaker~[\CodLev]).
And \PrevThm\ will go through if the metric is~$C^2$.  So, it would have been
sufficient to have assumed the manifold to be $C^3$ and the basic fields
on it to be $C^2$. (The manifold needs one higher order of
differentiability than the tensor fields on it because the transformation
formula for such fields when co-ordinates $x$ are changed to co-ordinates
$x'$ involves
$\partial x'\over\partial x$.)

The precise degree of differentiability, however, ought not
to matter in classical physics. Since the
matter and the geometrical fields, as well as the spacetime manifold
itself, are probably only approximations to more fundamental underlying
structures, we ought to be able to consider smooth enough
such approximations. Mathematically this can be made precise, for it
may be shown that $C^r$ structures may indeed be approximated by
$C^{\infty}$ ones
\StoreRef{\MunkDT}
\Ref{J. Munkres, \Jou{Elementary Differential Topology}, Princeton
University Press, Princeton (1966).}.
For these reasons, it seems justified to work
always with smooth quantities (and not have to keep track at each
stage of the number of derivatives).

This justification for assuming that the quantities that we are considering
are smooth, works if they were differentiable to begin with. What would
happen if they were only continuous? If these quantities are fields defined
on a given manifold, then they too can be smoothed. But, if we are
considering mappings from one manifold to another (as we are, between
the manifolds ${\cal S}_1$ and ${\cal S}_2$, in this paper), then continuous
mappings cannot always be smoothed. This occurs in situations involving the
so-called `exotic'
differentiable manifolds~[\MilnStash\,
\StoreRef{\UhlFr}
\Refm{K. Uhlenbeck and D. Freed, \Jou{Instantons and Four-Manifolds},
Springer Verlag, New York (1984).}
\StoreRef{\Witten}
\Refe{E. Witten, \Jou{Comm. in Math. Phys.}, \Vol{100}, 197 (1985).}.
There are arguments by Witten~[\Witten]
that such structures might be interesting to consider. To explain what is
involved, let ${\cal M}_1$ and ${\cal M}_2$ be two smooth manifolds of
the same dimension and let $f$ be a 1-1 mapping of ${\cal M}_1$ onto
${\cal M}_2$. $f$ is called a {\it homeomorphism\/} if it and $f^{-1}$
are continuous, and it is called a {\it diffeomorphism\/} if it and
$f^{-1}$ are differentiable. If a diffeomorphism exists and is
$C^r$, then a $C^{\infty}$ diffeomorphism also exists~[\MunkDT], and so
it is not important for many purposes
to specify the degree of differentiability of~$f$. It is important to
note that when we talk about two manifolds being diffeomorphic or
not, we are not making a statement about the manifolds only as point
sets or as topological spaces. Our statement refers also to the
{\it differential structure\/} on the manifolds, i.e., the maximal class of
coordinate systems on each manifold that are compatible with each other
(in the sense
that when two coordinate systems overlap, the transformations from one to
the other are differentiable (smooth, if the manifold is to be called
smooth)). It turns out to be
possible to define on the same topological space two
different differential structures, i.e., to have manifolds that are
homeomorphic, but not diffeomorphic. A differentiable manifold
that is homeomorphic to some standard differentiable manifold like $R^n$ or
$S^n$, but not diffeomorphic to it, is called an {\it exotic differentiable
manifold}. Examples of this exist for $S^n$, $n\geq 7$~[\MilnStash\,
\Refe{J. Milnor, in \Jou{Lectures in Modern Mathematics, II}, edited by
T.L. Saaty, Wiley, New York (1964).}
and for $R^4$~[\UhlFr].

How is all this related to ``topology change''? If we are to consider
exotic differentiable manifolds, then a natural question to ask is if
they can be created from ordinary ones, i.e., is it possible to find
an interpolating spacetime with boundaries that are homeomorphic
but not diffeomorphic? Would the theorems of this paper apply to such a
situation?

First consider the question at the manifold level. Here, examples of
interpolating manifolds exist. For example, there is an
8-dimensional manifold (which is constructed by looking at vectors of
length $\leq 1$ in a certain type of $R^4$ bundle over $S^4$)
whose boundary may be
shown to be an exotic $S^7$~[\MilnStash]. In this manifold remove a ball
of radius~$\epsilon$ around some point~$p$. The resulting manifold will
be a cobordism between an ordinary and an exotic~$S^7$. Next, can a
Lorentz metric be put on this cobordism? Since it is 8-dimensional,
it can be modified so that its Euler characteristic vanishes. On this
cobordism the required vector field $V^a$ will exist, and so the
cobordism will be Lorentzian. The theorems of this paper should then
apply. Thus, though the creation of exotic manifolds in
relativity appears to be kinematically possible, this process, too,
appears to be
dynamically forbidden within a classical framework of the
type being considered here.

It is worth observing that this conclusion depends on $V^a$ being
differentiable. If we were satisfied with metrics that are only
continuous (as we might be, if we were interested purely in causal structure
and not in dynamics), then the associated $V^a$ would also only be
continuous. In this case, ${\cal S}_1$ and ${\cal S}_2$ need not be
diffeomorphic, even if causality violations are forbidden. Indeed,
${\cal S}_1$ and ${\cal S}_2$ need not even be homeomorphic here,
since the integral curves of $V^a$ need not be unique~[\CodLev].

\Section Concluding Comments

Since part of the point of this paper is to address (and, with luck, dispel)
certain misconceptions about Lorentzian topology change, here is a list of
a few of the more common ones, along with some comments:

\Bull{Topology change is intrinsically incompatible with a Lorentzian
metric}.
Much of section~III addresses this and shows that
this perception is not true.

\Bull{Two-dimensional topology change is necessarily singular}.
This perception appears to be based on the studies that have
been made~[\ADeW\,\MCD]
of the $S^1\cup S^1\to S^1$ transition (the so-called ``trousers topology'').
In this case there is a singularity, but the examples of section~III
show that there are also non-singular topology-changing
spacetimes in two dimensions (albeit with closed timelike curves).

\Bull{Closed-universe topology change leads to
closed timelike curves only when the metric is time-orientable}.
This is addressed in section~VII. As shown there, it is possible to
use non-time-orientable metrics to avoid closed
timelike curves only when one of the boundaries is empty.

\Bull{Closed-universe topology change leads to
closed timelike curves only if some suitable energy condition holds}.
Neither Geroch's original theorem, nor its mild generalization in
section~IV, assume anything about the energy-momentum tensor,
or indeed about a field equation~-- the results are purely kinematical.

\Bull{Closed-universe topology change leads either to closed timelike
curves or to a singularity}.
The truth of this depends on the
definition of a singularity. If the standard incomplete-geodesic
definition is used, then this statement is not true:
as long as the causal compactness condition is met, causality
violations have to occur when the topology changes, even if incomplete
geodesics are admitted. But,
other definitions of a singularity
may make the statement true: this is briefly
discussed at the very end of this paper.

\Bull{Topology change may be dynamically had in closed universes if the
metric is allowed to be singular}.
The comments under the previous misconception apply here as well.
The situation here is complicated, however, by the existence of a
further theorem due to Tipler~[\Tipl] that was originally presented~--
and has been quoted~-- as a singularity theorem. Compactness assumptions
are not made in this theorem, and topology change is then shown to lead
to singularities, {\it but only under a significant additional
assumption}. The result is discussed further below.

Of course, all of these comments are valid only under the assumptions
of this paper. It is best not to view them too dogmatically: it is always
possible that different conclusions may be drawn if different assumptions
are made.

The same cautionary note applies to the main theorems of this
paper which, following on the earlier work of Geroch and Tipler, appear to
forbid topology change. Their true value is not so
much that they actually rule out topology change, but rather that they
allow us to pinpoint what modifications we have to make in our general
framework so as to allow it. A popular modification (for a
number of reasons, not all directly related to the specific problem
of topology change) is to abandon the Lorentzian framework altogether
and to use a Euclidean path integral formalism. But even within the
general Lorentzian framework there are still several interesting
possibilities.

\SubSection Dropping causal compactness

One possibility is to drop the causal compactness assumption. But,
it is hard to see that it could be replaced by a weaker assumption
that still restricts the candidates for interpolating manifold in
some way. And without some restriction, as we have seen, these
manifolds can be cut and truncated in an entirely arbitrary manner.
Also, there is the theorem of Tipler (ref.~[\Tipl], Theorem~5)
that was mentioned above: this result shows (in the closed universe
case) that if topology change occurs via a non-compact interpolating
spacetime, then it contains (under some mild additional
assumptions) a singularity (in the
sense of an incomplete timelike geodesic) {\it or a point at infinity}.
That it is a singularity that must occur may be inferred only under the
significant additional assumption that there is an upper bound on the
lengths of certain timelike curves in the region of interest. It
has been shown by Yodzis~[\Yodz] that we have a certain amount of choice in
the matter: a conformal transformation may be found in some cases to make the
interpolating spacetime future complete (i.e., the singularity may
be pushed to infinity). But the presence of points at infinity is still
a highly undesirable feature. It seems reasonable, therefore, to
retain some type of compactness assumption.

\SubSection Weakening the curvature constraints

We might also consider weakening the constraints on the curvature
in Theorem~3. This would not affect the presence of causality
violations, but it might permit topology change as a dynamical
process. A drastic step in this direction would be to alter
Einstein's equation so that assumptions~(i) and~(ii) can no longer
be justified from reasonable restrictions on the source.
But such an alteration would have to be fairly severe~-- for Einstein's
equation was used only in a very weak way in going from a condition
on $T_{ab}$ to a condition on $R_{ab}$. There is no real
justification~-- theoretical or
experimental~-- for such a step. Another possibility is (as
was discussed earlier) that there might be violations of the energy
conditions large enough to allow assumption~(ii) to be violated.
Some discussions of wormhole creation are, for example,
based precisely on large violations of the energy condition
(see ref.~[\MTY] and other references cited therein). The other
assumption is fairly benign. It might be interesting to try and weaken
it even more, but this is unlikely to allow topology change. One
possible weakening would replace ``every full null geodesic'' in
the statement by ``almost every full null geodesic''. It has been
argued by Sorkin
\Ref{R. Sorkin, unpublished remark.}
that such a statement would more truly be
a `generic condition'. Singularity
theorems have been proved with this kind of weaker generic condition
\Ref{A. Borde, \Jou{J. of Math. Phys.}, \Vol{28}, 2683 (1987).}
and it would be worth trying to do the same here.

\SubSection Degenerate metrics

Finally, we return to the interesting idea~-- discussed by
Sorkin~[\SorkTwo], Ashtekar~[\Asht], Horowitz~[\Horow], and others,
and mentioned briefly in the Introduction~-- that it might be possible
to allow the metric to vanish or to become degenerate at isolated points
and to use these kinds of singularities in order to get topology change.
Now, one problem with allowing the metric to vanish is that we cannot
compute its inverse and so cannot calculate the curvature at the points
where it vanishes. It appears, however, that the new spinorial variables
introduced by Ashtekar
\Ref{A. Ashtekar, \Jou{Phys. Rev. Lett.}, \Vol{57}, 2244 (1986);
\Jou{Phys. Rev.}, \Vol{D36}, 1587 (1987).}
to describe relativity might allow
this to be done. In this approach we can formulate all the basic
equations without having to `raise and lower indices', i.e., without
having to use the inverse of the metric.
Also, even in standard general relativity (couched in first-order
language) Horowitz~[\Horow] has shown that it is possible to
construct reasonable topology-changing spacetimes if degenerate metrics
are allowed. So, this might well prove to
be the correct approach to describing topology change.

It is worth pointing out here that
there is a further problem with degenerate (or vanishing) metrics:
the causal structure
that is normally associated with a Lorentz metric will not necessarily
be well-defined. But there is a way around this that allows in some cases
a definition of causal relationships,
even at points where the metric behaves badly.
This will be discussed in detail elsewhere
\Ref{A. Borde and R. Sorkin, in preparation.}.

\Sectionvar Acknowledgements

It is a pleasure to thank Rafael Sorkin for sparking my interest
in topology change and for several extremely helpful comments
on the manuscript. I also thank him,
Luca Bombelli and Tom Roman for many stimulating discussions on
the topics discussed here, and Matt Visser for some helpful
comments on an early version of this paper.
Financial
support was provided by NSF grants PHY8318350 and PHY8310041 to the
Relativity Group of Syracuse University in the initial stages of this work
and by a grant from Long Island University during the final stages.
The final version of the paper was written when I was a guest,
first of the High Energy Theory Group of Brookhaven National Laboratory,
and then of the Institute of Cosmology at Tufts University;
I thank both institutions for their hospitality.

\EndPaper